\documentclass[11pt,a4paper]{article}
\pdfoutput=1
\usepackage{jcappub}
\usepackage{bm}
\usepackage{enumitem}
\usepackage{braket}
\usepackage{ulem}
\def\dd{\mathrm{d}}
\def\mcA{\mathcal{A}}

\def\mcP{\mathcal{P}}

\def\inf{{\rm inf}}
\def\osc{{\rm osc}}
\def\BD{{\rm BD}}
\def\tot{{\rm tot}}

\def\Mpl{M_{\rm Pl}}
\def\GeV{{\rm GeV}}

\def\G{{\rm G}}
\def\reh{{\rm reh}}
\def\mbx{\bm{x}}
\def\mbk{\bm{k}}
\def\tmax{\text{max}}
\def\tmin{\text{min}}
\def\EM{\text{EM}}
\def\GW{\text{GW}}
\def\peak{\rm peak}
\def\Mpc{{\rm Mpc}}

        \def\fin{{\rm fin}}

\def\osc{{\rm osc}}

\def\0{{(0)}}
\def\sig0{\dot{\sigma}_0}

\def\ph0{\dot{\phi}_0}

\title{Chiral Gravitational Waves Produced in a Helical Magnetogenesis Model}
\author[a]{So Okano}
\author[b]{and Tomohiro Fujita}

\affiliation[a]{Department of Physics, Tokyo Institute of Technology,
2-12-1 Ookayama, Meguro-ku, Tokyo 152-8551, Japan
}
\affiliation[b]{Institute for Cosmic Ray Research, University of Tokyo, Kashiwa, 277-8582, Japan}

\emailAdd{okano.s.ab@m.titech.ac.jp}
\emailAdd{tfujita@icrr.u-tokyo.ac.jp}

\abstract{We investigate the gravitational wave production induced by the primordial magnetic fields in a parity-violating magnetogenesis model.
It is shown that the gravitational waves detectable by LISA, DECIGO or BBO 
and the magnetic fields strong enough to explain the blazar observation can be simultaneously produced. 
The magnetic fields and the gravitational waves have the same chirality and their amplitudes are related, which may also be tested by future observations.
}

\arxivnumber{2005.13833}

\begin{document}

\maketitle


\section{Introduction}
It is well known that galaxies and their clusters have magnetic fields with the typical strength  $\mathcal{O}(10^{-6}\G)$~\cite{Bernet:2008qp,Bonafede:2010xg,Feretti:2012vk}. However we still do not know where they come from.  Recent  multi-frequency blazar observations imply the existence of the magnetic fields in the void region and the lower bound of such intergalactic magnetic fields (IGMFs) as~\cite{Neronov:1900zz,Dolag:2010ni,Essey:2010nd,Tavecchio:2010ja,Taylor:2011bn,Vovk:2011aa,Takahashi:2013lba,Chen:2014rsa,Finke:2015ona,Biteau:2018tmv}
\begin{align}
\label{eq:observation bound}
B_{\rm eff}\,\gtrsim\, 10^{-16}{\rm G},
\qquad
B_{\rm eff} \equiv B\times\left\{\begin{array}{cc}
1 & (\lambda\geq 1\Mpc)\\
\sqrt{\lambda/1\Mpc} \quad& (\lambda\leq 1\Mpc)
\end{array}\right.,
\end{align} 
where $B$ and $\lambda$ are the strength and the correlation length of the IGMFs, respectively. In addition, the observations of cosmic microwave background (CMB) give the upper bound to the large scale magnetic fields, $B\lesssim \mathcal{O}(n\G)$ for $\lambda \gtrsim 1$Mpc~\cite{Ade:2015cva}. For more details, interested readers are referred to review articles~\cite{Durrer:2013pga,Subramanian:2015lua}

The generation mechanism of the magnetic fields of galaxies and their clusters are  divided into the astrophysical and cosmological scenarios. The former includes Biermann battery effect~\cite{Hanayama:2005hd} in which the non-parallel gradient of electron pressure and density play important rolls to generate  small-scale magnetic fields. However it is difficult to explain the large-scale magnetic fields and IGMFs, because there are not much astrophysical activities in the void region. The latter, the cosmological scenario hypothesizes that primordial magnetic fields (PMFs) are formed in the early universe before the recombination, and they are amplified to the order of $10^{-6}\G$ via the galactic dynamo which is driven by the interaction between magnetic fields and plasma~\cite{Davis:1999bt}. This scenario can explain the existence of IGMFs. However, we do not know the origin of such PMFs. One of the candidates is the cosmological phase transition~\cite{Vachaspati:1991nm,Enqvist:1993np,Grasso:1997nx,Ellis:2019tjf} in which bubbles from the first order phase transition of the universe serve the kinetic energy to the electromagnetic fields via the bubble collisions. Another candidate is the Harrison mechanism~\cite{Takahashi:2005nd,Saga:2015bna}, in which the second order perturbations of the electron, proton, and photon induce the electric current and it becomes the source of the magnetic fields. Inflationary magnetogenesis, in which the quantum fluctuation is the origin of the PMFs, is one of the most studied scenario. Since the standard $U(1)$ gauge fields on the flat-FLRW universe can not be amplified by inflation due to its conformal symmetry~\cite{Benevides:2018mwx}, several models are devised to break the conformal symmetry during inflation.

The kinetic coupling model~\cite{Ratra:1991bn,Bamba:2003av} was first proposed by Ratra~\cite{Ratra:1991bn}, where a rolling scalar field $\Phi$ coupled to the electromagnetic fields as $I^2(\Phi)F_{\mu\nu}F^{\mu\nu}$ and the electromagnetic fields are generated while $I(\Phi)$ evolves. Unfortunately, the original model inevitably violates one of the following conditions~\cite{Demozzi:2009fu,Fujita:2012rb,Fujita:2013qxa,Fujita:2014sna}; (i) the effective coupling constant between the canonical electromagnetic fields and charged particles should be small enough to validate the perturbative treatment, (ii) the energy density of the electromagnetic fields should not overwhelm the inflaton energy density during inflation, and (iii) the curvature perturbation induced by the generated magnetic fields should be consistent with the CMB observations. Since these three conditions are often broken in the models aiming to explain Eq.~\eqref{eq:observation bound} in the literature, they are known as the serious problems to achieve viable inflationary magnetogenesis and called (i) the strong coupling problem, (ii) the back reaction problem, and (iii) the curvature perturbation problem, respectively. 
Recent studies have proposed to  introduce the IR cut-off into the spectrum of the produced magnetic fields~\cite{Ferreira:2013sqa,Ferreira:2014hma} and the post-inflationary phase of magnetogenesis~\cite{Kobayashi:2014sga,Fujita:2016qab,Vilchinskii:2017qul} to satisfy these conditions. Another well studied model of inflationary magnetogenesis is the axial coupling model~\cite{Turner:1987bw,Garretson:1992vt,Field:1998hi,Anber:2006xt,Fujita:2015iga} first proposed by Turner and Widrow~\cite{Turner:1987bw}, and its 
detailed analytic study was provided by Anber and Sorbo~\cite{Anber:2006xt}. In this model, a rolling pseudo scalar serves its kinetic energy to the electromagnetic fields via the axial coupling $\phi F_{\mu\nu}\tilde{F}^{\mu\nu}$. Ref.~\cite{Fujita:2015iga} numerically found that the significant amplification of the magnetic fields occurs around the end of the inflation and Ref.~\cite{Adshead:2016iae} showed that the generated magnetic fields reach to $10^{-16}\G$ by the lattes simulation.

Caprini and Sorbo~\cite{Caprini:2014mja,Caprini:2017vnn}  proposed the hybrid model which contains both the kinetic and axial couplings, $I^2(\chi)(FF-\gamma F\tilde{F})$, where $\chi$ represents a rolling pseudo scalar. Since this model violates the parity symmetry, the generated magnetic fields have non-zero helicity. Then the correlation length of the helical magnetic fields in the plasma grows faster than the cosmic expansion via the inverse cascade~\cite{Son:1998my,Christensson:2000sp,Kahniashvili:2012uj}, which is a well-known mechanism in magnetohydrodynamics (MHD). By using the inverse cascade, they show that this model can produce the helical magentic fields with the strength $B_{\rm eff}=\mathcal{O}(10^{-17}\G)$. This model was extended by including magnetogenesis during the reheating era, and the produced magnetic fields in Ref.~\cite{Sharma:2018kgs,Fujita:2019pmi} can be even larger.
In addition, some inflationary magnetogenesis models, not included in the above categories, also achieve to generate magnetic fields consistent with Eq.~\eqref{eq:observation bound} ~\cite{Domenech:2015zzi, Mukohyama:2016npi,Brandenburg:2020vwp}.

Though we have the indirect observational implication of the PMFs and the successful magnetogenesis models, the PMFs are not yet well established. Thus it is important to seek the other observables associated with the PMFs. The stochastic gravitational waves background is one of the most important observables to reveal the primordial universe, since they can directly bring us the information of the universe before the recombination. Since the PMFs have large intensity right after the generation, they can be the primary source of the gravitational waves (GWs). Several researches have argued that the nature of the GWs induced by the PMFs highly depend on its magnetogenesis model~\cite{Sorbo:2011rz,Caprini:2014mja,Namba:2015gja,Jimenez:2017cdr,Saga:2018ont,Sharma:2019jtb,Ozsoy:2020ccy}. Therefore  magnetogenesis models can be distinguished by the observation of the GWs. 
In this paper, we study the GW production in the magnetogenesis model proposed in Ref.~\cite{Fujita:2019pmi}. This model can generate strong and helical electromagnetic fields by considering their amplification during not only inflation but also the subsequent reheating era. Therefore, there is a good chance to produce primordial GWs with large amplitudes. Indeed, we will show that chiral GWs are obtained from the generated magnetic fields, and the strength of the GWs can be sufficiently large to be observed by LISA and DECIGO.

This paper is organized as follows. In section \ref{sec:Model}, we give the brief review on the hybrid magnetogenesis model proposed in Ref.~\cite{Fujita:2019pmi}. The mechanism to generate the GWs from the magnetic fields, the estimation of their power spectrum and the comparison between our numerical results and the sensitivity of interferometers are given in section \ref{sec:GW production}. Finally, section \ref{sec:summary} is devoted to the summary and discussion.

\section{Helical Magnetogenesis Model}
\label{sec:Model}

Here we briefly review an inflationary magnetogenesis model first proposed in Ref.~\cite{Caprini:2014mja} and further developed in Ref.~\cite{Fujita:2019pmi}. 

\subsection{Model setup}

Let us consider the following Lagrangian proposed in Ref.~\cite{Caprini:2014mja}  on a spatially flat FLRW metric $\dd s^2=a^2(\eta)\left[-\dd \eta^2+\dd \mathbf{x}^2\right]$ with the conformal time $\eta$:
\begin{align}
\label{eq:action}
\mathcal{L}=&\dfrac{\Mpl^2}{2}R-\dfrac{1}{2}(\partial_{\mu}\phi)^2-V(\phi)-\dfrac{1}{2}(\partial_{\mu}\chi)^2-U(\chi)-\dfrac{1}{4}I^2(\chi)\left(F_{\mu\nu}F^{\mu\nu}-\gamma F_{\mu\nu}\tilde{F}^{\mu\nu}\right).
\end{align}
Here $\Mpl$ is the reduced Planck mass, $R$ is the Ricci scalar, $\phi$ is the inflaton, $\chi$ is a spectator scalar field, $F_{\mu\nu}$ is the field strength of U(1) gauge field, $\tilde{F}^{\mu\nu}=\epsilon^{\mu\nu\alpha\beta}/(2\sqrt{-g})F_{\alpha\beta}$ is the dual field strength, and $\gamma$ is a constant. V($\phi$) and U($\chi$) are the potentials of $\phi, \chi$ respectively. We assume that $\chi$ has a non-zero vacuum expectation value and coupled to the U(1) gauge field via the kinetic coupling function $I(\chi)$ but its energy density is always much smaller than the total energy density. Thus, we call $\chi$ a ``spectator''. One can derive the equation of motion (EoM) for $\chi$ and see that the kinetic energy of $\chi$ is transferred to the electromagnetic fields through the time derivative of $I(\chi)$. Since electromagnetic fields are produced while $I(\chi)$ evolves in time, for simplicity, we treat $I$ as a function of time by ignoring the perturbation of $\chi$. We also assume that $I(\eta)$ varies during inflation and the subsequent reheating era (i.e. the inflaton oscillation era).
In order to study the generation of the gauge field, we promote the classical vector potential to the quantum operator as
\begin{align}
\label{eq:U(1) quantization}
A_i(\eta,\mbx)=\sum_{\lambda=\pm}\int\dfrac{d^3k}{(2\pi)^3}e^{i\mbk\cdot\mbx}e^{(\lambda)}_i(\hat{\mbk})\left[a^{(\lambda)}_{\mbk}\mcA_{\lambda}(\eta,k)+a^{(\lambda)\dagger}_{-\mbk}\mcA_{\lambda}^{\ast}(\eta,k)\right],
\end{align}
where we work in Coulomb gauge, $A_0=\partial_iA_i=0$, 
$e^{(\pm)}_i(\hat{\mbk})$ are the right/left-handed polarization vectors which satisfy $i\epsilon_{ilm}k_le^{(\pm)}_m(\hat{\mbk})=\pm k e^{(\pm)}_i(\hat{\mbk})$, $\mcA_{\lambda}$ is the mode function of the gauge field, and we impose the usual canonical commutation relation, $\left[a^{(\alpha)}_{\mbk},a^{(\beta)\dagger}_{-\mbk'}\right]=(2\pi)^3\delta(\mbk+\mbk')\delta^{\alpha\beta}$. Here $\hat{\bm{n}}$ represents the unit vector parallel to $\bm{n}$.
The EoM for the mode function is written
as\begin{align}
\label{eq:gauge mode function}
\left[\partial_{\eta}^2+k^2\pm2\gamma k\dfrac{\partial_{\eta} I}{I}-\dfrac{\partial_{\eta}^2I}{I}\right]\Big(I(\eta)\mcA_{\pm}(\eta,k)\Big)=0.
\end{align}
Note that we restore the usual EoM in normal electromagnetism when the function $I$ is constant, and the gauge modes are not amplified. To solve Eq.~\eqref{eq:gauge mode function}, we need to specify $I$ as the function of the conformal time. 
\begin{figure}
        \centering
        \includegraphics[width=\textwidth]{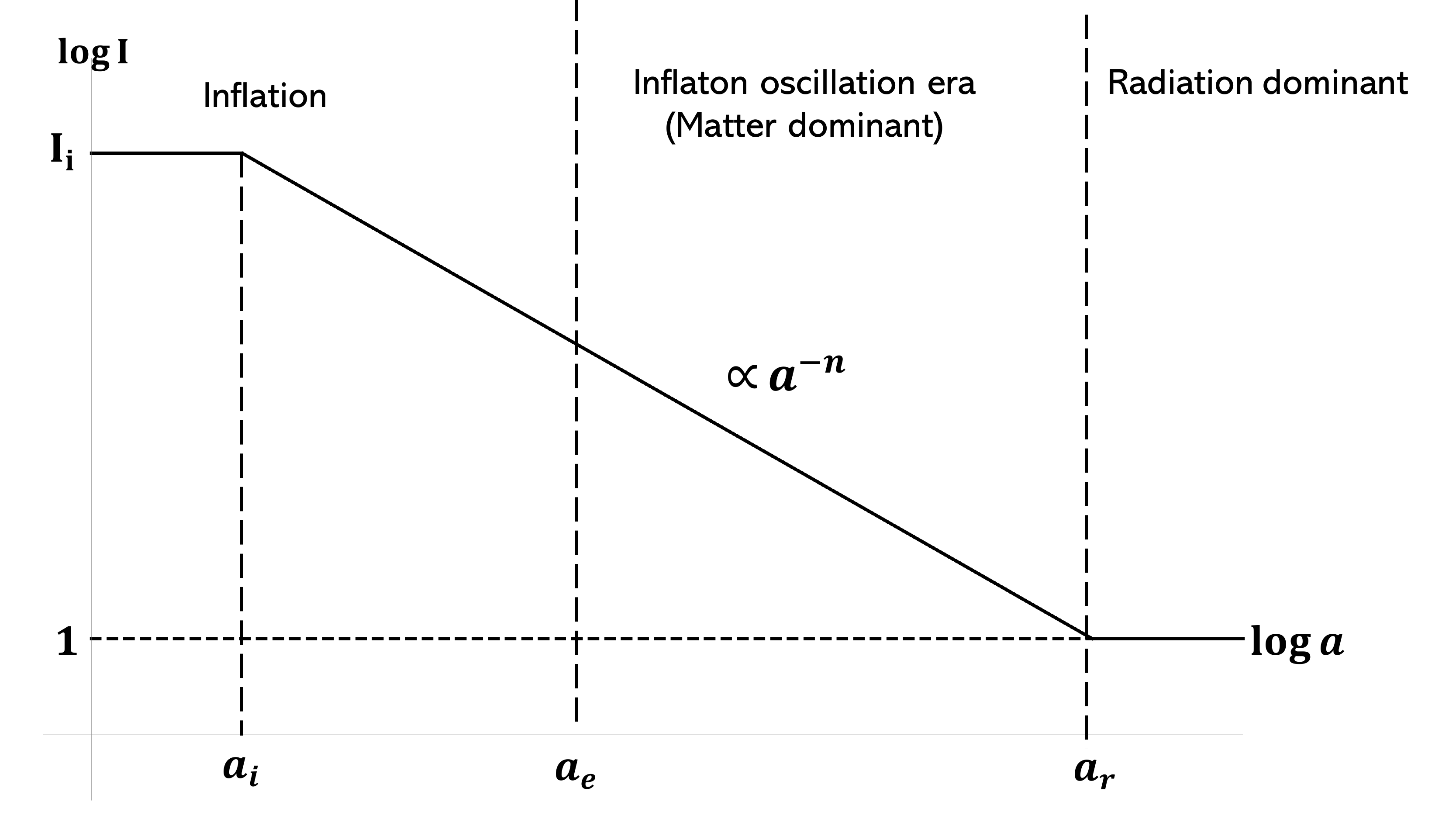}
        \caption{The behavior of $I(a)$ given in Eq.~\eqref{eq:I}. $I(a)$ decreases in proportion to $a^{-n}$ for $a_i<a<a_r$.  
In this paper, we assume that $I(a)$ stops  at the reheating completion, $a_r$.}
        \label{I:figure}
\end{figure}
In a  similar way to Ref.~\cite{Fujita:2019pmi},
we assume that the kinetic function $I$ is given by
\begin{align}
\label{eq:I}
I(\eta)=\left\{
\begin{array}{ll}
(a_i/a_r)^{-n}\equiv I_i & (a < a_i) \\
(a(\eta)/a_r)^{-n} &  (a_i<a<a_r) \\
1                  & (a_r<a) 
\end{array}\right.\,,
\end{align}
where $I(\eta)$ begins to vary at $a_i$ during inflation, and $I(\eta)$ becomes unity at the reheating completion $a_r$. 
The shape of the function $I(a)$ is illustrated in Fig.~\ref{I:figure}.
The conformal time in each era behaves as
\begin{align}
\label{eq:conformal time}
\eta=\left\{
\begin{array}{ll}
-1/aH_{\inf}\propto a^{-1} & (a < a_e) \\
2/aH\propto a^{1/2} &  (a_e<a<a_r) \\
1/aH\propto a                & (a_r<a) 
\end{array}\right.\,,
\end{align}
where inflation ends at $a_e$ and the reheating is completed at $a_r$.
By using Eq.~\eqref{eq:gauge mode function} and Eq.~\eqref{eq:I}, we obtain the EoM in each era as 
\begin{align}
\label{eq:onset}
\left[\partial_{\eta}^2+k^2\right]\left(I\mcA^{\BD}_{\pm}\right)&=0, \
(a<a_i)\\
\label{eq:inflation}
\left[\partial_{\eta}^2+k^2\pm
2\xi\dfrac{k}{\eta}-\dfrac{n(n+1)}{\eta^2}\right]\left(I\mcA^{\inf}_{\pm}\right)&=0,
\ (a_i<a<a_e)\\
\label{eq:oscillation}
\left[\partial_{\eta}^2+k^2\mp
4\xi\dfrac{k}{\eta}-\dfrac{2n(2n+1)}{\eta^2}\right]\left(I\mcA^{\osc}_{\pm}\right)&=0,
\ (a_e<a<a_r)\\
\label{eq:stopped}
\left[\partial_{\eta}^2+k^2\right]\left(I\mcA^{\fin}_{\pm}\right)&=0, \
(a_r<a),
\end{align}
where $\xi\equiv n\gamma$. The term including $\xi$ generates the polarization of the gauge fields through the tachyonic instability which is effective around the horizon crossing of a fluctuation mode, $k\eta \sim \xi$. Since  we are interested in the efficient production of the polarized gauge field, we assume that $\xi >n ~ \rm{(i.e. \gamma>1)}$. 
In addition, the super-horizon mode is amplified by the term  proportional to $\eta^{-2}$. These amplifications occur while the function $I(a)$ varying.

\subsection{Solving the dynamics of the electromagnetic fields}

Now we can solve the EoM,   Eqs.~\eqref{eq:onset}-\eqref{eq:stopped} by using Bunch-Davies initial conditions and the junction conditions at the boundary between different eras. From Bunch-Davies initial conditions,  the solutions for $a<a_i$ is written as $I_i\mcA^{\BD}_{\pm}=e^{-ik(\eta-\eta_i)}/\sqrt{2k}$. The gauge fields and their time derivatives are continuous at any time and hence we impose the junction conditions, $\mcA_{\pm}(\eta_*-\delta)=\mcA_{\pm}(\eta_*+\delta)$ and 
$\partial_{\eta}\mcA_{\pm}(\eta_*-\delta)=\partial_{\eta}\mcA_{\pm}(\eta_*+\delta)$
in the limit $\delta\to 0$ at the boundary times $\eta_*=\eta_i$ and $\eta_e$.
The solution during inflation is already given in Ref.~\cite{Fujita:2019pmi}. In this paper, we are interested in the solution during reheating, and we show only the solution for Eq.~\eqref{eq:oscillation}.

The solution for Eq.~\eqref{eq:oscillation} can be written as
\begin{align}
\label{eq:oscillation sol}
I\mcA^{\rm reh}_{\pm}=\dfrac{1}{\sqrt{2k}}\left[D_{1}^{\pm}(|k\eta_e|)\,M_{\pm 2i\xi,2n+\frac{1}{2}}(2ik\eta)+D_{2}^{\pm}(|k\eta_e|)\,W_{\pm 2i\xi,2n+\frac{1}{2}}(2ik\eta)\right],
\end{align}
where $W_{\alpha,\beta}(z)$ and $M_{\alpha,\beta}(z)$ are Whittaker functions, and the coefficients $D_1^{\pm}, D_{2}^{\pm}$ which depend on the wave number of the mode through $|k\eta_e|$ are given by
\begin{align}
D_1^{\pm}(y)
=&\frac{1}{4iy}\frac{\Gamma(2n+1\mp 2i\xi)}{\Gamma(4n+2)}C_2^{\pm}(k\eta_i)
\Big[W_{1\pm 2i\xi,2n+1/2}(4iy)W_{\mp i\xi,n-1/2}(-2iy)
\notag\\&\qquad\qquad\qquad\qquad\qquad\qquad\quad-2W_{1\mp i\xi,n-1/2}(-2iy)W_{\pm 2i\xi,2n+1/2}(4iy)\Big],\\
D_2^{\pm}(y)
=&\frac{1}{4iy}\frac{\Gamma(2n+1\mp 2i\xi)}{\Gamma(4n+2)}C_2^{\pm}(k\eta_i)
\Big[2M_{\pm 2i\xi,2n+1/2}(4iy)W_{1\mp i\xi,n-1/2}(-2iy)
\notag\\
&\qquad\qquad\qquad\quad-(1+2n\pm 2i\xi)2W_{1\pm i\xi,2n+1/2}(4iy)W_{1\mp i\xi,n-1/2}(-2iy)\Big].
\end{align}
 Here, we introduced $y\equiv |k\eta_e|$
and the function $C_{2}^{\pm}$ is 
\begin{align}
C_2^{\pm}(k\eta_i)
=\dfrac{\Gamma(n\pm i\xi)}{2ik\eta_i\Gamma(2n)}\left[(n\mp i\xi)M_{1\mp i\xi,n-\frac{1}{2}}(2ik\eta_i)
-(n\mp i\xi-2ik\eta_i)M_{\mp i\xi,n-\frac{1}{2}}(2ik\eta_i)\right].
\end{align}
In the sub-horizon limit at the onset of the $I(\eta)$ evolution, $-k\eta_i\gg1$, one can use $C_2^{\pm}\simeq e^{\pm\frac{1}{2}\pi\xi}$.\ On the other hand, in the super-horizon limit, $-k\eta_i\ll1$, one obtains $C_2^{\pm}\simeq\Gamma(n\pm i\xi)|2ik\eta_i|^n/\Gamma(2n)$.
One finds that  in Eq.~\eqref{eq:oscillation sol}, the term including
$D_1^\pm$ corresponds to the growing mode and the other is the decaying mode, and the growing mode is always dominant component of the mode function.
For the super-horizon modes at the end of inflation, $|k\eta_e|\ll 1$, $D_1$ is approximated as,
\begin{align}
\label{eq:D1 sup}
D_1^{\pm}(y)
\approx\dfrac{2^{-5n}\Gamma(2n)}{(4n+1)\Gamma(n\pm i\xi)}y^{-3n}C_2^{\pm}
\qquad (|k\eta_e|\ll1).
\end{align}
The mode function at the super-horizon scale is written as,
\begin{align}
\label{eq:mode function sup}
I\mcA^{\reh}_{\pm}&\simeq\dfrac{1}{\sqrt{2k}}D_1^{\pm}(|k\eta_e|)\,
(2k\eta)^{2n+\frac{3}{2}} \qquad (k\eta\ll1),
\end{align}
where we used the super-horizon approximation of the Whittaker function, $M_{a,b}(x)\simeq x^{b+1/2}$ for $x\ll1$. In this case, one can safely use Eq.~\eqref{eq:D1 sup} since $|k\eta_e|<k\eta\ll1$. On the other hand, the mode function at the sub-horizon scale is obtained as
\begin{align}
\label{eq:mode function sub}
IA^{\reh}_{\pm}\simeq \frac{1}{\sqrt{2k}}D_1^{\pm}(|k\eta_e|)
\frac{e^{\pi\xi}\Gamma(4n+3)}{\Gamma(2n+\frac{3}{2}\mp 2i\xi)}e^{ik\eta}(k\eta)^{-2i\xi} \qquad (k\eta\gg 1).
\end{align}
Here we used the sub-horizon approximation of the Whittaker function, $M_{a,b}(x)\simeq x^{-a}e^{x/2}\Gamma(2b+1)/\Gamma(b-a+1/2)$ for $x\gg1$. In this case, one cannot always use the super-horizon approximation for $D_1$.

\subsection{Electromagnetic Power spectra}

\begin{figure}
        \centering
        \includegraphics[width=.45\textwidth]{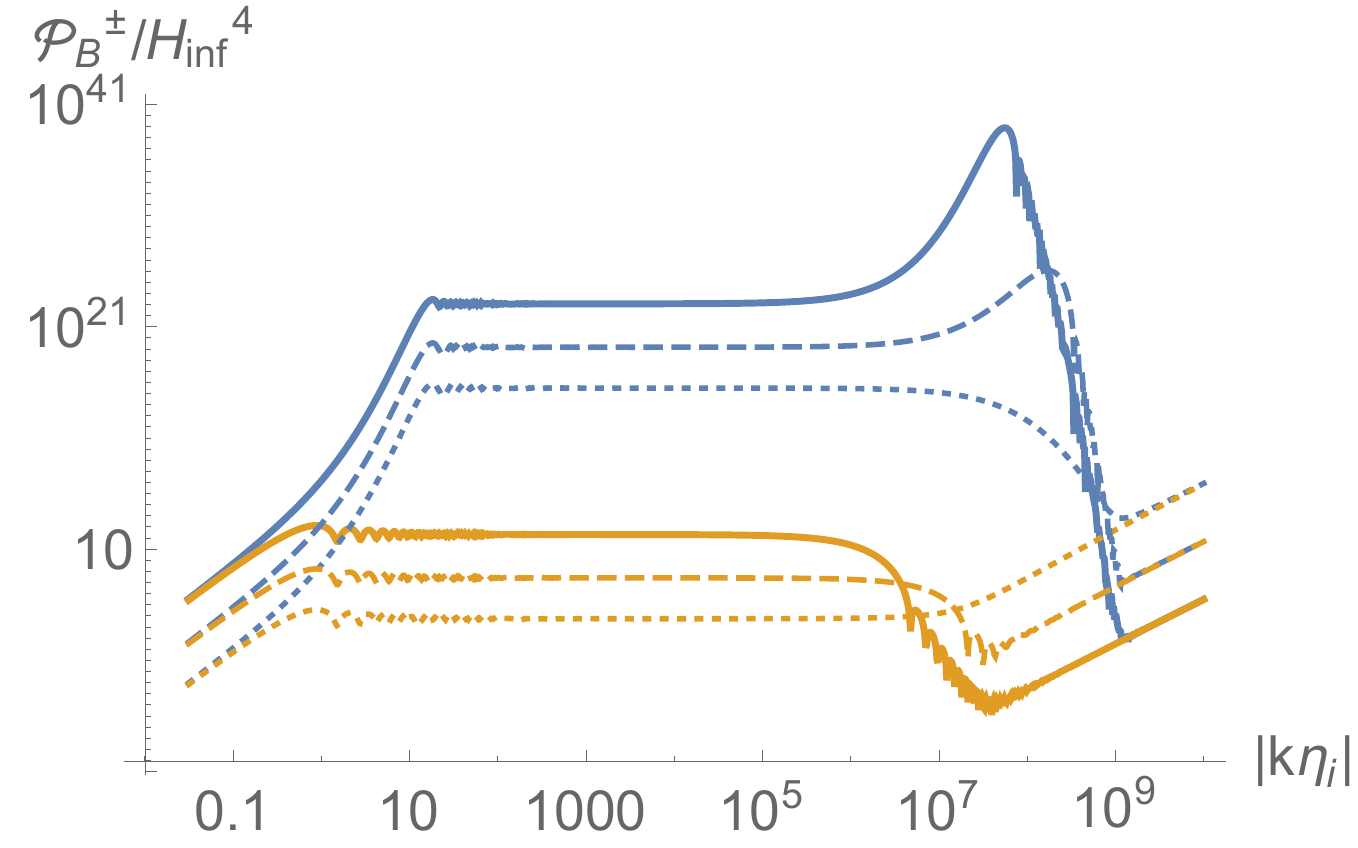}
        \includegraphics[width=.45\textwidth]{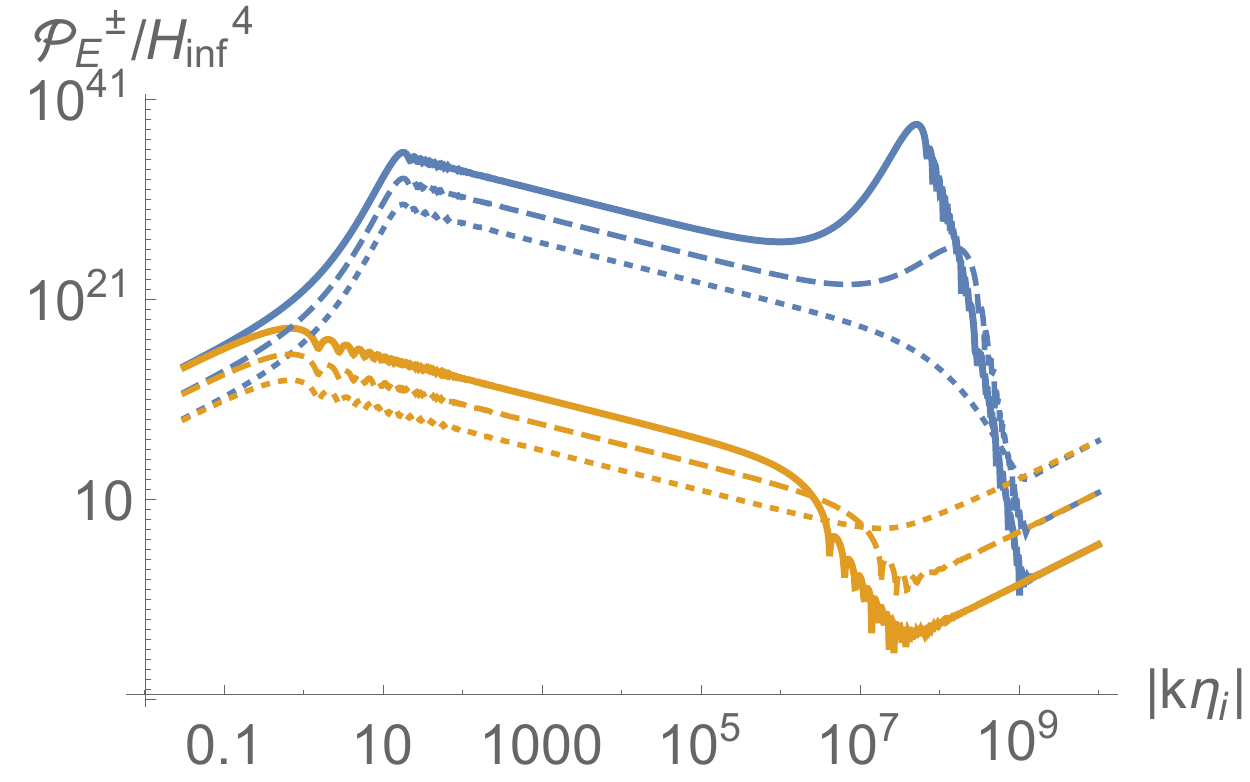}
\caption{The magnetic and the electric power spectra are shown in the left and right panels, respectively. The dotted, dashed and solid lines denote the power spectra at $a=a_e,\, 20a_e$ and $400a_e$ respectively. 
We fixed the parameters as $n=3,\, \xi=7.6$ and $a_e=7.7\times 10^7a_i$.
The right-handed modes (blue) are amplified by the tachyonic instabilities, while the left-handed modes (orange) are not, because we chose $\xi>0$.
 }
        \label{fig:powerspectrum of EB}
\end{figure}

\begin{figure}
        \centering
        \includegraphics[width=.45\textwidth]{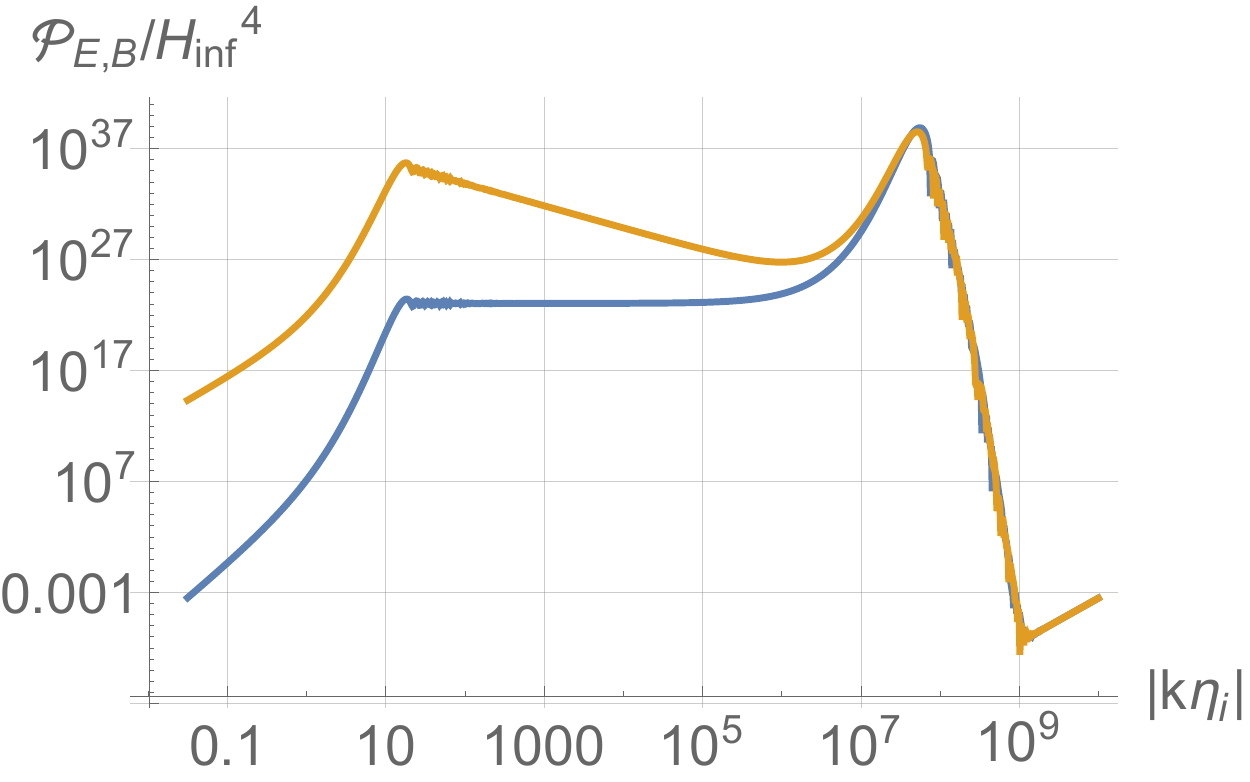}
        \includegraphics[width=.45\textwidth]{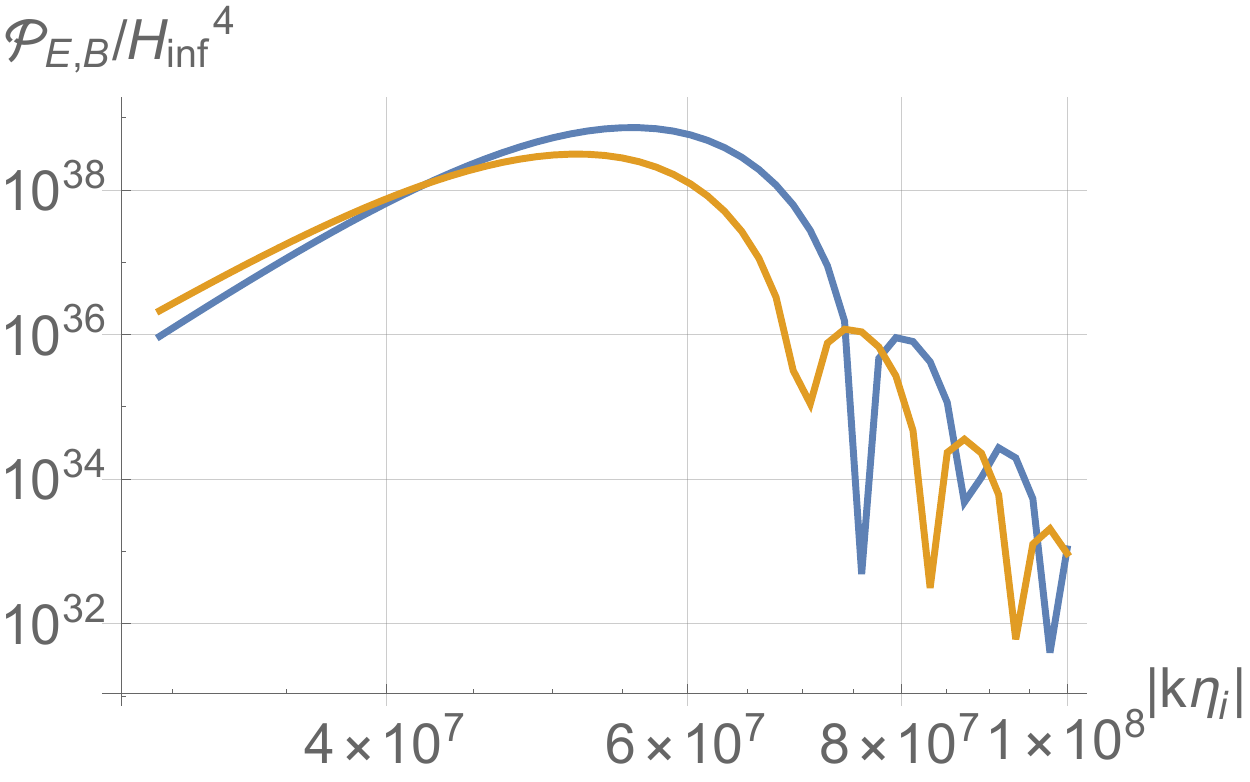}
        \caption{
The magnetic (blue) and electric (orange) power spectra are compared at $a=10^2a_e$. The right panel is the enlarged figure of the left panel around the small scale peak, $k=4\xi/\eta$. 
The magnetic fields are stronger than the electric fields around the small scale peak, while the electric fields are stronger than the magnetic fields on larger scales. The parameters are the same as Fig.~\ref{fig:powerspectrum of EB}.}
        \label{fig:comparison of EB}
\end{figure}

By using the analytic solutions of $\mcA_{\pm}^{\reh}(\eta)$, we can calculate the power spectra of the electromagnetic fields. We define the power spectra of the electric and magnetic fields as
 \begin{align}
 \label{eq:def ps}
 \mcP_{E}^{\pm}(\eta,k)\equiv\dfrac{k^3I^2}{2\pi^2a^4}\left|\partial_{\eta}\mcA_{\pm}\right|^2, 
\qquad\quad
\mcP_{B}^{\pm}(\eta,k)\equiv\dfrac{k^5I^2}{2\pi^2a^4}\left|\mcA_{\pm}\right|^2.
 \end{align}
In Figs.~\ref{fig:powerspectrum of EB} and \ref{fig:comparison of EB}, 
we show the magnetic and electric power spectra for the both circular polarizations, and compare them. 
One can see that the electric and magnetic fields are maximally helical and they have peaks. 
Both the magnetic and electric power spectra have sharp peaks on the same scale which is relatively small, and only the electric power spectrum
has a peak on a larger scale at $k \sim |\eta_i|^{-1}$.
Let us scrutinize the origin of these small scale peaks.
The  tachyonic instability  during inflation occurs slightly before the modes exit the horizon $k\sim -2\xi/\eta\propto a$, and only the right-handed modes are amplified. On super-horizon scales all the modes are increased at the same rate due to the last term in Eq.~\eqref{eq:inflation}, and the smooth spectra are formed for $-1/\eta_i\lesssim k\lesssim -2\xi/\eta$ as shown as the dotted lines in Fig.~\ref{fig:powerspectrum of EB}. On the other hand, the second tachyonic amplification during the subsequent 
reheating era takes place slightly after the modes re-enter the horizon $k\sim 4\xi/\eta\propto a^{-1/2}$. Thus the modes in $4\xi/\eta<k<2\xi/|\eta_e|$  acquire the double tachyonic amplifications. 
Since the modes quickly decay well inside the horizon, the mode which has just gone through the second tachyonic amplification has the largest amplitude.
As a result, the power spectra obtain such sharp peaks at $ k=4\xi/\eta$.
While the electric fields are stronger than the magnetic fields at the large scale, the magnetic fields are stronger than the electric fields at the peak scale as shown in the  Fig.~\ref{fig:comparison of EB}.
One analytically confirms these behaviors by using super/sub horizon approximation for the mode function. 
Note that the small scale peaks are located at $k\eta=4\xi$ which is well inside
the horizon for $4\xi\gg 1$.
The ratio of the power spectra is estimated as,
 \begin{align}
 \label{eq:comparison of the power spectra}
 \frac{\mcP_B}{\mcP_E}(\eta_r)=k^2\frac{\left|\mcA_{\pm}\right|^2}{\left|\partial_{\eta}\mcA_{\pm}\right|^2}\sim\left\{\begin{array}{cc}
 4 & (k\eta_r=4\xi)\\
 \dfrac{4k\eta^2}{(3+4n)^2} \ll 1& (k\eta _r\ll 1)
 \end{array}\right..
 \end{align}
Here we used Eqs.~\eqref{eq:mode function sup} and \eqref{eq:mode function sub}. From now on, following the previous paper \cite{Fujita:2019pmi}, we use these fiducial values of the model parameters,
\begin{equation}
n=3, \qquad \xi=7.6.
\label{parameter choice}
\end{equation}

\subsection{Consistency conditions}
\label{Consistency conditions}

For consistent magnetogenesis, there should not be the significant back reaction from the electromagnetic fields to the background evolution of the universe
which was assumed in Eq.~\eqref{eq:conformal time}. Then the maximum value of the energy fraction of the electromagnetic fields is much less than unity,
\begin{align}
\label{eq:energy fraction}
\Omega_{\EM}\equiv \frac{1}{2\rho_{\tot}}\int \frac{dk}{k}\left(\mcP_{E}(k)+\mcP_{B}(k)\right)\ll 1,
\end{align}
where $\rho_{\tot}=3\Mpl^2H^2$ is the total energy density of the universe. Since the power spectra of the electromagnetic fields have prominent peaks 
at two different scales,
we separate the main contributions to $\Omega_{\EM}$ as,
\begin{align}
\Omega_{\EM}(\eta_e<\eta\le\eta_r)\simeq \left. \Omega_{\EM}\right|_{k\eta\sim 4\xi}+ \left.\Omega_{\EM}\right|_{|k\eta_i|\sim2\xi}.
\end{align}
Here $\left. \Omega_{\EM}\right|_{k\eta\sim 4\xi}$ denotes the energy fraction contributed from the small scale peaks, and $\left.\Omega_{\EM}\right|_{|k\eta_i|\sim2\xi}$  comes from the large scale peak. 
$\Omega_{\rm EM}(\eta)$ takes its maximum value when the electromagnetic fields stop growing.
Thus we impose the back reaction condition on the energy fraction at $\eta=\eta_r$ as
\begin{align}
\label{eq:constraint}
\left.\Omega_{\EM}(\eta_r)\right|_{k\eta\sim 4\xi}=10^{-2}, 
\qquad 
\left.\Omega_{\EM}(\eta_r)\right|_{|k\eta_i|\sim2\xi}\ll10^{-2},
\end{align}
where we require that the large scale peak has a negligible contribution
compared to the small scale peaks. In the following, we evaluate these contributions to determine the model parameters which satisfy the above conditions.

\begin{figure}
        \centering
        \includegraphics[width=.8\textwidth]{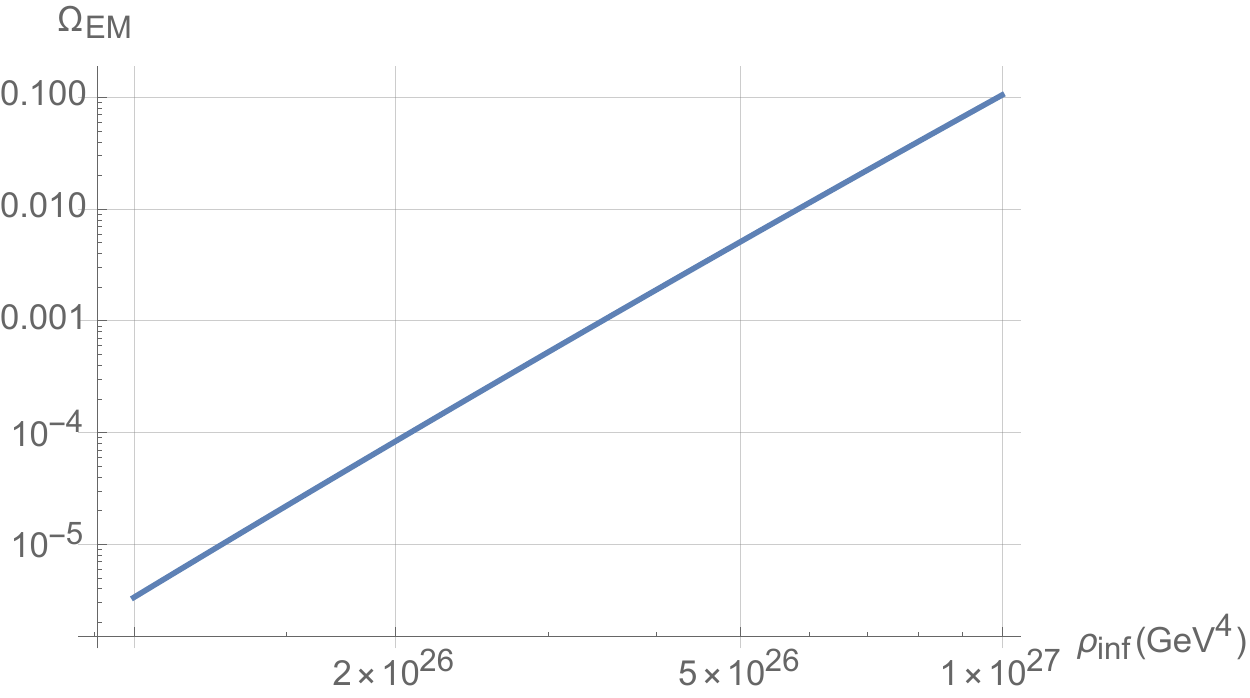}
        \caption{The relation between $\Omega_{\EM}$ and $\rho_{\inf}$ for $T_r=2.2\times 10^4\GeV$. To satisfy the consistency condition, $\Omega_{\EM}\ll1$, we choose $\rho_{\inf}<5.85\times 10^{26}\GeV^4$. There is the lower bound on $\rho_{\inf}$ since the energy density at reheating completion is smaller than $\rho_{\inf}$, $\rho_{\inf}>\pi^2g_{\ast}T_r^4/30\approx 10^{18}\GeV^4$. }
        \label{fig:omegaEM}
\end{figure}

Let us evaluate the energy fraction from the small scale peaks at $k\eta=4\xi$. Since the magnetic fields are four times stronger than the electric fields at the peak scale (see Eq.~\eqref{eq:comparison of the power spectra}), the right-handed magnetic part is the main component of $\left.\Omega_{\EM}(\eta_r)\right|_{k\eta\sim 4\xi}$. Hence we evaluate the following integral,
\begin{align}
\label{eq:EM fraction}
\left.\Omega_{\EM}(\eta_c)\right|_{k\eta\sim 4\xi}&\simeq
\frac{1}{2\rho_{\tot}}\int \frac{dk}{k}\mcP_{B}(k)\,
\notag\\
&\simeq \dfrac{1}{96\pi^2\Mpl^2a_r^2\eta_r^2}\int_{1}^{4\xi}\frac{dy}{y}y^4\left|D_{1}^{+}\left(\frac{1}{2}e^{-\frac{N_r}{2}}y\right)M_{ 2i\xi,2n+\frac{1}{2}}(2iy)\right|^2, 
\end{align}
where we used the $H=2/a\eta$ at the matter dominant era and we introduced the dummy variable $y\equiv k\eta_r$. Since the significant contributions to $\left.\Omega_{\EM}(\eta_c)\right|_{k\eta\sim 4\xi}$ are  in sub-horizon scale, we ignore the super-horizon modes, $0<k\eta_r<1$. Additionally, the modes on  smaller scales than the peak scale, $k>4\xi/\eta_r$, have negligible contributions, because they have smaller amplitudes and are highly oscillating. Therefore the interval of the integration of $k\eta_r$  can be limited into $1<k\eta_r<4\xi$. $N_r$ represents the e-folding number between the end of inflation and that of the reheating phase,
\begin{align}
N_r&=\ln\left(\frac{\eta_r}{2|\eta_e|}\right)^2=\ln\left(\frac{a_r}{a_e}\right)=\frac{1}{3}\ln\left(\frac{\rho_{\inf}}{\frac{\pi^2}{30}g_{\ast}T_r^4}\right)\,
\notag\\
&\approx 8.05+\frac{4}{3}\ln\left(\frac{\rho_{\inf}^{1/4}}{10^{7}\GeV}\right)-\frac{4}{3}\ln\left(\frac{T_r}{10^{4}\GeV}\right)-\frac{1}{3}\ln\left(\frac{g_{\ast}}{100}\right),
\end{align}
where $T_r$ and $g_{\ast}$ denote  the temperature and the number of degree of freedom at the reheating completion, respectively. By using the equation of the entropy conservation we obtain the condition for $a_r$ as
\begin{align}
a_r\approx8.0\times 10^{-18}\left(\frac{T_r}{10^{4}\GeV}\right)^{-1}\left(\frac{g_{ \ast s}}{100}\right)^{-\frac{1}{3}},
\end{align}
where we introduce the number of degree of freedom for entropy, $g_{\ast s}$, and we assume that $g_{\ast s}=g_{\ast}=100$ from now on. $\eta_r$  is also represented by $T_r$ as,
\begin{align}
\eta_r=2e^{\frac{1}{2}N_r-N_{i}}|\eta_i|\approx 1.18 \times 10^{-11} \left(\frac{T_r}{10^4\GeV}\right)^{-1}\Mpc,
\label{etar and Tr}
\end{align}
where $N_i\equiv\ln(a_e/a_i)\approx 29.9+2/3\ln(\rho_{\inf}/10^7\GeV)+1/3\ln(T_r/10^4\GeV)-\ln(k_i/1\Mpc^{-1})$. One can numerically evaluates Eq.~\eqref{eq:EM fraction} by substituting  particular values into $T_r$ and $\rho_{\inf}$. We show the relation between $\rho_{\inf}$ and $\Omega_{\EM}$ in Fig.~\ref{fig:omegaEM} for a fixed reheating temperature, $T_r=2.2\times 10^4\GeV$. 
In this case, one finds that $\rho_{\inf}\approx6\times 10^{26}\GeV^4$ satisfies our consistency condition, Eq.~\eqref{eq:constraint}.

Let us consider the large scale contribution to $\Omega_{\EM}$. 
The large scale peak can be larger than the small scale peak,  since the electric power spectrum depends on $k^{-2}$ at super-horizon scales for $n=3$. 
We evaluate the contribution of the large scale electric fields 
\begin{align}
\label{eq:EM fractionsh}
\left.\Omega_{\EM}(\eta_r)\right|_{k\eta_i|\sim2\xi}&\approx\frac{1}{2\rho_{\tot}}\int \frac{dk}{k}\mcP_{E}(k\eta\ll 1),
\notag\\
&=\frac{25}{2\pi^3}\dfrac{\rho_{\inf}}{\Mpl^4}e^{2N_i+5N_r}\dfrac{\mathcal{I}(\xi)\sinh(\pi\xi)}{\xi^5+5\xi^3+4\xi},
\notag\\
&\approx 10^{-3}\left(\frac{T_r}{10^{4}\GeV}\right)^{-6}\left(\frac{\rho_{\inf}^{1/4}}{10^{7}\GeV}\right)^{3}\left(\frac{k_i}{10^7\Mpc^{-1}}\right)^{-2}.
\end{align}
Here we introduce the numerical fits $\mathcal{I}(\xi)\equiv\int_0^{\infty}\frac{d(-k\eta_i)}{(-k\eta_i)^3}|C_2^{+}(-k\eta_i)|\approx\frac{e^{\pi\xi}}{4.7\xi^2+2.7\xi+10}$ valid for $1\leq\xi\leq 50$~\cite{Fujita:2019pmi}. Then, one can suppress the contribution from the super-horizon mode by sending $k_i$ to a sufficiently small scale.

\subsection{Inverse cascade and the present magnetic field strength}

Here we evaluate the present magnetic field strength by considering  the inverse cascade process in magnetohydrodynamics. It is known that when helical magnetic fields and plasma tightly interact with each other in the turbulent and high conductivity regime, the correlation length of the magnetic field increases, 
because of the magnetic helicity 
conservation. This process is called the inverse cascade. Our magnetogenesis scenario generates maximally helical, small scale and strong magnetic fields. Thus
the subsequent inverse cascade automatically works and enables us to obtain  magnetic fields
whose correlation length is much larger than the case only with the adiabatic expansion. This property is advantageous to explain the observational lower bound, Eq.~\eqref{eq:observation bound}.

The magnetic helicity density $\mathcal{H}(\eta)$ is defined as the volume average of the local helicity $\bm{A}\cdot\bm{B}(\eta,\bm{x})$,
 \begin{align}
 \label{eq:helicity density}
 \mathcal{H} &\equiv\frac{1}{V}\int_V \dd^3x\  \epsilon_{ijk}A_i\partial_j A_k\,
 \\
 &\simeq  a^3 \lambda_{\rm phys}B_{\rm phys}^2,
 \qquad (\rm maximally\ helical)
 \label{eq:H simple}
 \end{align}
where $V$ is the comoving volume, $\bm{B}_{\rm phys}\equiv -a^{-2} \bm{\nabla}\times\bm{A}
\simeq a^{-1}\bm{A}/\lambda_{\rm phys}$ is the physical magnetic field
and $\lambda_{\rm phys}$ is its physical correlation length.
The magnetic field on the boundary of the volume $V$ is assumed to have no normal component for the gauge
invariance.
It is known that the helicity is conserved in the magnetohydrodynamics limit in which the electric conductivity becomes infinite. In the context of cosmology, the helicity approximately conserves after the reheating completion.
It can be also shown that the helicity represents the difference between the right handed and left handed polarization contributions
to Eq.~\eqref{eq:helicity density}. In our case, the right handed component is much stronger than left handed one  and the generated magnetic fields are maximally helical.
Then the simplified evaluation, Eq.~\eqref{eq:H simple}, is available.

Now let us consider the inverse cascade process for our maximally helical magnetic fields.   Provided that  the helicity density $\mathcal{H}$ in Eq.~\eqref{eq:helicity density} is  conserved, the helicity at the end of the magnetogenesis and the present time are the same. $B_{\rm phys}$ and $\lambda_{\rm phys}$ can be evaluated by the maximum value of the  magnetic power spectrum 
$\mathcal{P}_B(k_{\peak})$ and $2\pi a/k_{\peak}$, respectively. Thus the helicity evaluated at $\eta=\eta_r$ is given by
 \begin{align}
 \mathcal{H} \simeq\  a^3_s\lambda_{\rm phys}(\eta_r)B_{\rm phys}^2(\eta_r)\simeq a_r^3\frac{2\pi a_r\eta_r}{4\xi}\mcP_{B}(\eta_r,k_{\peak}),
 \end{align}
where we used $k_{\peak}\equiv 4\xi/\eta_r$, 
$I^2B^2_{\rm phys}\,\simeq\,\mcP_{B}(k_{\peak})$
and $I(\eta_r)=1$.  On the other hand, the blazar observations are sensitive to $B_{\rm eff}$ defined in Eq.~\eqref{eq:observation bound}
which is directly related to the helicity at the present time as $B_{\rm eff}\simeq \sqrt{\mathcal{H}/1{\rm Mpc}}$ for $\lambda<1\Mpc$.
Therefore, we obtain $B_{\rm eff}$ as,
 \begin{align}
 \label{eq:beffective}
B_{\rm eff}\approx
8\times 10^{-14}\G\left(\dfrac{\Omega_{\EM}}{10^{-2}}\right)^{1/2}\left(\dfrac{T_r}{ 10^4\GeV}\right)^{-1/2}\left(\dfrac{\xi}{10}\right)^{-1/2}.
\end{align}
Here we used $\Mpl=2.43\times10^{18}\GeV $ and $\G=6.8\times 10^{-20}\GeV^2$, $\Mpc=1.56\times 10^{38}\GeV^{-1}$,
and $\mcP_{B}(\eta_r,k_{\rm peak})\approx 2\rho_{\tot}\Omega_{\EM}$.
Hence, the observational constraint Eq.~\eqref{eq:observation bound}
can be explained in our model.

\section{Gravitational Wave Production}
\label{sec:GW production}

In this section, we show that U(1) gauge fields source 
gravitational waves (GWs) by considering the second order perturbation, and calculate the power spectrum of GWs. 
We are mainly interested in the small scale GWs which can be observed by the GW interferometers. 
Since we found that the magnetic fields are stronger than electric fields around the small scale peak in the previous section, we only consider the magnetic component in this section.

\subsection{U(1) gauge fields sourcing GWs}

Here we derive the EoM for the tensor perturbation with the U(1) source term. We introduce the perturbation of metric
around the FLRW background universe as
\begin{align}
\dd s^2=a^2(\eta)\left[-\dd \eta^2+\left(\delta_{ij}+h_{ij}\right)\dd x^i \dd x^j\right],
\end{align}
where $h_{ij}$ denotes the metric tensor perturbation with the transverse and traceless conditions $h^i_{\ i}=0, \partial_i h_{ij}=0$. 
From the Lagrangian Eq.~\eqref{eq:action}, one can derive the EoM for the tensor perturbation with the source term,
\begin{multline}
\label{eq:EoM for GW}
\tilde{h}_{\bm{k}}^{''(\lambda)}(\eta)+2aH\tilde{h}_{\bm{k}}^{'(\lambda)}(\eta)+k^2\tilde{h}_{\bm{k}}^{(\lambda)}(\eta)
\\=-\dfrac{2I^2(\eta)}{M^2_{pl}a^2}e_{ij}^{(\lambda)}(-\bm{k})\epsilon_{ikl}\epsilon_{jnm}
\int\dfrac{\dd^3q}{(2\pi)^3}\,q_k (k_n-q_n) 
A_l(\eta,\bm{q})A_m(\eta,\bm{k}-\bm{q}),
\end{multline}
where  $e^{(\lambda)}_{ij}(\mbk)$ is the polarization tensor.
In  eq.~\eqref{eq:EoM for GW}, we ignored the contribution from the electric fields and took Fourier transformation, 
\begin{align}
h_{ij}(\eta,\bm{x})=\int \dd^3k\,\tilde{h}_{ij}(\eta,\bm{k})\, e^{i\bm k\cdot \bm x}.
\end{align}
By using the Green function $G_k(\eta,\eta')$ for the tensor perturbation in the matter dominant era,
\begin{align}
G_{k}(\eta,\eta')=\eta'\theta(\eta-\eta')
\dfrac{\left(k^2\eta\eta' +1\right) \sin (k (\eta -\eta'
        ))+k (\eta' -\eta ) \cos (k (\eta -\eta' ))}{k^3\eta ^3},
\end{align}
where $\theta(\eta-\eta')$ denotes the Heaviside step function,
we obtain the solutions for Eq.~(\ref{eq:EoM for GW}) as
\begin{align}
\tilde{h}_{\bm{k}}^{(\lambda)}(\eta)=-\frac{2}{\Mpl^2}e_{ij}^{(\lambda)}(-\bm{k})
&\epsilon_{ikl}\epsilon_{jnm}
\int\dfrac{\dd^3q}{(2\pi)^3}q_k (k_n-q_n)
\notag\\\times&\int \dd\eta'G_k(\eta,\eta')\dfrac{I^2(\eta')}{a^2(\eta')}
A_l(\eta',\bm{q})A_m(\eta',\bm{k}-\bm{q}).
\end{align}
With the quantized gauge fields, Eq.~(\ref{eq:U(1) quantization}), the two point function of the sourced gravitational waves is given by
\begin{align}
\label{eq:power spectrum}
\Braket{\tilde{h}_{\bm{k}}^{(\lambda)}\tilde{h}_{\bm{k}'}^{(\lambda)}(\eta)}
&=\delta(\bm k+\bm k')\,\dfrac{1}{2\Mpl^4}\sum_{\alpha,\beta=\pm}\int
\dd^3p\  p^2|\mbk+\bm{p}|^2 \left(1-\alpha\lambda\hat{\bm{p}}\cdot\hat{\bm{k}}\right)^2\left(1+\beta\lambda(\widehat{\bm k+\bm p})\cdot\hat{\bm{k}}\right)^2
\notag\\
& \ \ \ \ \ \ \ \ \ \ \ \ \ \ \ \ \ \ \ \ \ \ \ \ \ \ \ \quad\times
\left|\int
\dd\eta'\, G_{k}(\eta,\eta')\dfrac{I^2(\eta')}{a^2(\eta')}\mcA_{\alpha}(\eta',p)\mcA_{\beta}(\eta',|\bm{k}+\bm{p}|)\right|^2,
\notag\\
&=\dfrac{(2\pi)^5}{2k^3}\delta(\bm k+\bm k')\mcP_{\lambda}(\eta,k),
\end{align}
where $\mcP_{\lambda}(\eta,k)$ is the dimensionless power spectrum of the induced GWs with the circular polarization label $\lambda$.
One can find that the peak of the GWs is located at $k=8\xi/\eta$ because the source term, the right hand side of Eq.~\eqref{eq:power spectrum}, is represented as the convolution of two gauge fields, and the integrand becomes biggest when the momenta of the both convoluted mode functions are $p=|\bm{k}+\bm{p}|=4\xi/\eta$. In this case, $\mbk$ and $\bm{p}$ satisfy the relation, $\mbk+\bm{p}=-\bm{p}$. Then the peak scale of the GWs is evaluated as $k^{\peak}_{\GW}=2k^{\peak}_{\EM}=8\xi/\eta_r$.

We will show the numerical calculation of the GWs power spectrum in section~\ref{sec:numerical calculation}. Before that, we analytically make the order estimate of 
the power spectrum around the peak, $k^{\peak}_{\GW}=8\xi/\eta$ as
\begin{align}
\label{eq:estimation of power spectra}
\mcP_{\GW}(\eta_r,k_{\GW}^{\peak})&\approx 
\frac{(k_{\GW}^{\peak})^7}{(2\pi)^5\Mpl^4}\int \frac{dp}{k_{\EM}^{\peak}}(k_{\EM}^{\peak})^3(4\pi)\left|\int d\eta'\dfrac{2\pi^2a^2G_{k_{\GW}^{\peak}}(\eta_r,\eta')}{(k_{\EM}^{\peak})^{5}}\dfrac{(k_{\EM}^{\peak})^{5}}{2\pi^2a^4}(I\mcA_{+}(\eta',\mbk_{\EM}^{\peak}))^2\right|^2 \notag\\
&\approx\dfrac{2^{2}\times a_r^4}{(k_{\GW}^{\peak})^{4}(k_{\GW}^{\peak}\eta_r)^8\Mpl^4}
\mcP_B^2(k_{\rm EM}^{\rm peak})\,
|\chi(k_{\GW}^{\peak} \eta_r)|^2\notag\\
&\approx 
 9\xi^{-4} \Omega_{\EM}^2,
\end{align}
where we used $\int\frac{\dd p}{p}\mcP_B(p)\approx \mcP_B(k_{\rm EM}^{\rm peak})\approx 2\rho_{\rm tot}\Omega_{\rm EM}$, and introduced 
$\chi$ as,
\begin{align}
\chi(z)&\equiv\int dz'\frac{\theta(z-z')z'^5\left(\left(z z' +1\right)\sin(z -z'))+(z' -z)\cos(z -z' )\right)}{z^3}\approx z^4\, \, \, (z\geq\mathcal{O}(10)).
\end{align}
To compare our result with the sensitivity curves of observational equipments, we transform the dimensionless power spectrum in radiation dominant era to the energy fraction of the GWs per logarithmic interval of the wave number $k$
at present time~\cite{Nakayama:2008wy},
\begin{align}
\label{eq:transfer function}
\Omega_{\GW}(k,\eta_0)=\dfrac{1}{12}\left(\dfrac{k}{a_0H_0}\right)^2\left(\dfrac{\Omega_{m}}{\Omega_{\Lambda}}\right)^2\left(\dfrac{g_{\ast}(T_{in})}{g_{\ast 0}}\right)\left(\dfrac{g_{\ast s0}}{g_{\ast s}(T_{in})}\right)^{\frac{4}{3}}\left(\dfrac{\overline{3j_1(k\eta_0)}}{k\eta_0}\right)^2T_1^2\left(\frac{k}{k_{\rm eq}}\right)\mcP_{\lambda}(\eta_r,k),
\end{align}
where $\Omega_{\rm GW} \equiv \rho_{\tot}^{-1}\,\dd \rho_{\rm GW}/\dd \ln k$, the subscript 0 indicates that the quantity is evaluated at the present time, $\Omega_m/\Omega_{\Lambda}$ is the ratio of the matter component and dark energy in the universe today, and $g_{\ast}(T_{\rm in})$ denotes the relativistic degrees of freedom for temperature $T_{\rm in}$ at which the corresponding mode re-enter the horizon. $j_1(x)=1/x(\sin x/x-\cos x)$ is the spherical Bessel function, which is the solution of the EoM for GWs without source during the matter dominant era, and the bar denotes the amplitude of an oscillating function. $T_1^2(k/k_{\rm eq})$ is called the transfer function which connects the GWs re-entering the horizon at radiation dominant era and at the matter dominant era. $k_{\rm eq}$ denotes the wave number of the mode which re-enters the horizon at the matter-radiation equality. The transfer function is calculated as
\begin{align}
T_1(x)=1+1.57x+3.42x^2.
\end{align}
We can approximately estimate the coefficient of $\mcP_\lambda$ in the right hand side of Eq.~\eqref{eq:transfer function} for $k/2\pi=0.1 \rm Hz$  as $10^{-6}$. Combining Eqs.~\eqref{eq:estimation of power spectra} and \eqref{eq:transfer function}, we obtain 
\begin{equation}
\Omega_{\rm GW} \approx 10^{-13}\left(\frac{\Omega_{\EM}(\eta_r)}{10^{-2}}\right)^2,
\end{equation}
where we used $\Omega_m=0.3, \Omega_{\Lambda}=0.7, H_0^{-1}=4.33\times 10^{3}\Mpc, h_0=0.7,k_{\rm eq}=7.1\times 10^{-2}\Omega_mh_0^2\Mpc^{-1},$ and $\eta_0=1.4\times 10^3 \Mpc$.

\subsection{Numerical calculation of $\Omega_{\rm GW}$}
\label{sec:numerical calculation}
In this section, we numerically compute the power spectrum of the sourced GWs and compare it with the sensitivity curves of the upcoming interferometers. Substituting Eq.~\eqref{eq:oscillation sol} into Eq.~\eqref{eq:power spectrum}, one obtains the formula to evaluate the power spectrum of the GWs at the end of the inflaton oscillating era, $\mcP_{\lambda}(\eta_r,k)$, sourced by magnetic fields  around the peak scale as
\begin{align}
\label{eq:numerical formula}
\mcP_{\lambda}(\eta_r,k)&\simeq-\left(\dfrac{a_iH_{\inf}}{\Mpl}\right)^4(k\eta_i)^4\dfrac{(k\eta_r)^2}{(2\pi)^4a_r^4}\sigma_{\alpha\beta\lambda}(k\eta_r).
\end{align}
Here, where $\sigma_{\alpha\beta\lambda}$ denotes
\begin{align}
\label{eq:horn ps}
&\sigma_{\alpha\beta\lambda}(k\eta)\equiv
\int_{1}^{\infty}dx\int_{0}^{1} dy \left(1+\alpha\lambda\dfrac{xy+1}{x+y}\right)^2\left(1+\beta\lambda\dfrac{1-xy}{x-y}\right)^2
\notag\\&\times
\left(\dfrac{x^2-y^2}{4}\right)^{2}\left|D^{\alpha}_1\left(\frac{x+y}{2}k\eta_e\right)\right|^2\left|D^{\beta}_1\left(\frac{x-y}{2}k\eta_e\right)\right|^2\notag\\
&\times\left|\int_{2|k\eta_e|}^{k\eta}\hspace{-0.2cm}dz\dfrac{\left(\left(k\eta z +1\right)\sin(k\eta-z)+(z-k\eta)\cos(k\eta-z)\right)}{z^3}
M_{2i\alpha\xi,2n+\frac{1}{2}}\left(i\frac{x+y}{2}2z\right)M_{2i\beta\xi,2n+\frac{1}{2}}\left(i\frac{x\hspace{-0.05cm}-\hspace{-0.05cm}y}{2}2z\right)\right|^2.
\end{align}
The direct numerical computation of the above equation is possible, while
it would be expensive. Thus, we analytically perform the time integral with the following approximation,
\begin{align}
\label{eq:tophat approximation}
M_{ 2i\xi,2n+\frac{1}{2}}(2ix)
\quad \to \quad
f_\text{tophat}(x)=a\theta(x-x_{\tmin})\theta(x_{\tmax}-x),
\end{align}
where $a, x_{\tmax}, x_{\tmin}$ are specified as the height and  FWHM  of first peak, respectively. In the present case with Eq.~\eqref{parameter choice}, these parameters are determined as $a=8\times 10^{22}, x_{\max}=3.8$, and $x_{\min}=3.1.$
By substituting Eq.~\eqref{eq:tophat approximation} into Eq.~\eqref{eq:horn ps},
we analytically perform the time integral with respect to $z$. After that, we numerically compute the integrations of $x$ and $y$ (see Appendix~\ref{sec:Appendix} for the detailed computation.). Since the generated magnetic fields are almost completely helical, one can ignore the left-handed mode and take $\alpha=\beta=+$.
Note that the left-handed GWs are much smaller than right-handed one, because the peak of the right-handed magnetic fields at $k_{\EM}^{\peak}=4\xi/\eta$ does not contribute to the left-handed GWs. One finds this feature in Eq.~\eqref{eq:power spectrum}.
 Thus we only consider $\Omega_{\GW}^{(+)}$ and $\sigma_{+++}$ from now on.
 
\begin{figure}
        \centering
        \includegraphics[width=1\textwidth]{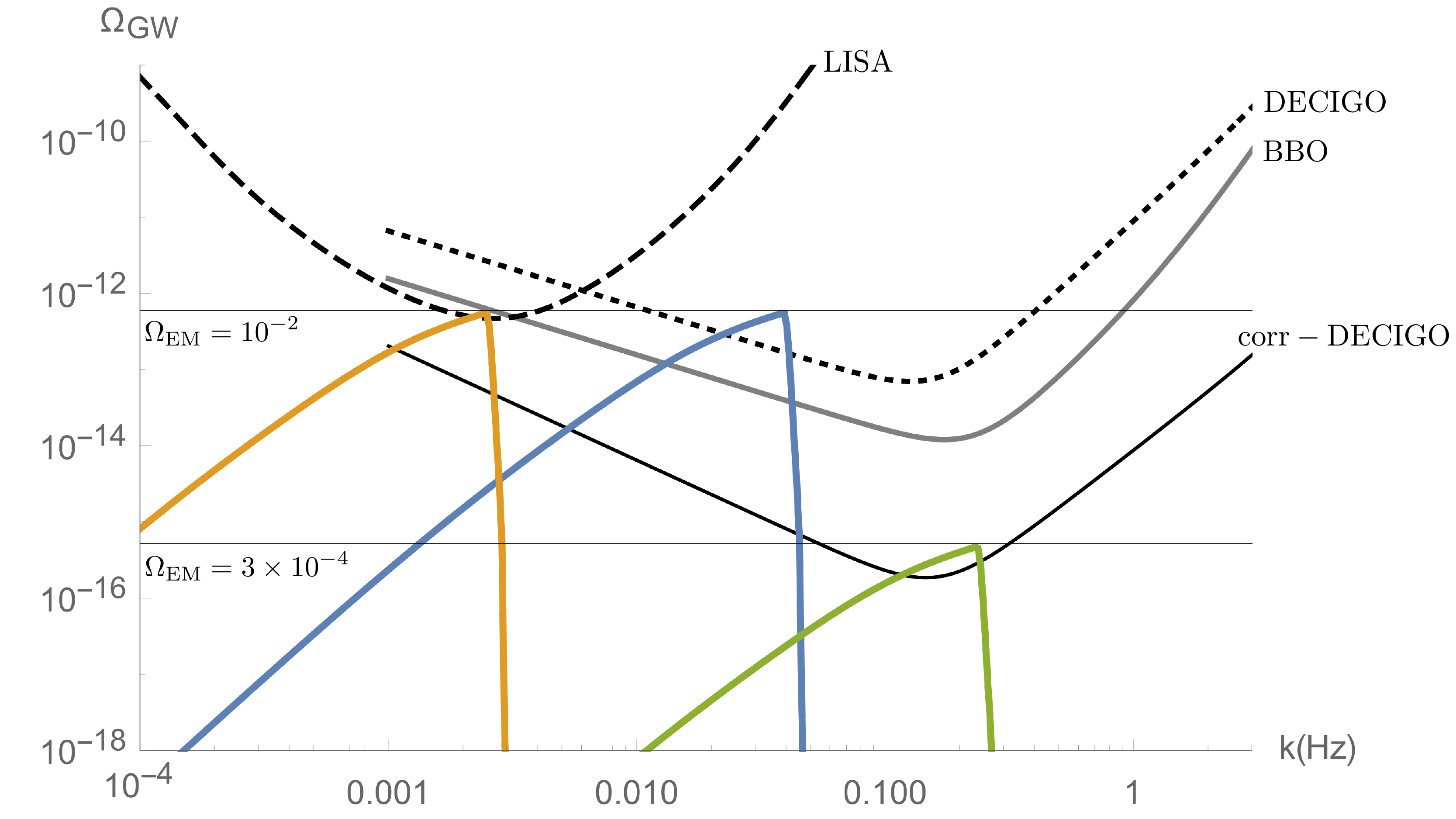}
        \caption{$\Omega_{\GW}(k)$s of the induced gravitational waves in our model are compared with the sensitivity curves of the future GW interferometers,
DECIGO (dotted), BBO (gray), LISA(dashed), and correlated DECIGO(thick). The thin gray lines denotes the peak amplitudes for given $\Omega_{\EM}$=$10^{-2}$, and $3\times 10^{-4}$, respectively. The parameters are fixed as $n=3, \xi=7.6, k_i=10^5\Mpc^{-1}$ for each $\Omega_{\GW}$ lines. The blue line is obtained by fixing $T_r=6.57\times10^{4}\GeV$, and $\rho_{\inf}^{1/4}=1.15\times 10^7\GeV$. The orange line is obtained by fixing $T_r=4.62\times10^{2}\GeV$, and $\rho_{\inf}^{1/4}=2.65\times 10^5\GeV$. The green line is obtained by fixing $T_r=1.10\times10^{4}\GeV$, and $\rho_{\inf}^{1/4}=2.36\times 10^6\GeV$. The maximum value of $\Omega_{\GW}(k)$ is calculated as $\Omega_{\GW}^{\max}=6\times 10^{-13}$. The peak scale can move between $10^{-8}{\rm Hz}<k_{\GW}^{\peak}<10^4{\rm Hz}$.}
        \label{fig:karidata}
\end{figure}

  In the rest of this section, we numerically perform $\Omega_{\GW}$ and compare it with the sensitivity curves of the GWs interferometers. First, we choose the fiducial parameters as $n=3$ and $\xi=7.6$ as stated in Eq.~\eqref{parameter choice}. 
We choose the reheating temperature $T_r$ to fix the peak scale $k^{\peak}_{\GW}=8\xi/\eta_r$
with Eq.~\eqref{etar and Tr} as
 \begin{align}
T_r=1.74\times 10^{5}\GeV\left(\frac{k^{\peak}_{\GW}}{1\rm Hz}\right).
 \end{align}
 In addition, one should determine the parameters $\rho_{\inf}$ and $k_i$ by using  Eqs.~\eqref{eq:constraint}, \eqref{eq:EM fraction}, and \eqref{eq:EM fractionsh}.
 Since the bottom of the sensitivity curves of DECIGO is located at $0.126\rm Hz$, for example, the peak scale is set to be the same value. In this case, the above equation fixes $T_r=2.2\times 10^4\GeV$. Then, Fig.~\ref{fig:omegaEM} is drawn and we find that  
 $\rho_{\inf}=5.85\times 10^{26} \GeV^4$ leads to  $\Omega_{\EM}=0.01$. Finally, we set $k_i=10^5 \Mpc^{-1}$ to satisfy Eq.~\eqref{eq:constraint}. 
 In general, we can tune the peak scale of the GWs by changing $T_r$ and then choose $\rho_{\inf}$ to achieve an arbitrary value of $\Omega_{\EM}$. 
 To satisfy the lower bound of IGMFs Eq.~\eqref{eq:observation bound} and the consistency for BBN, the parameter region of the reheating temperature is constrained as $10^{-2}<T_r/1\GeV<10^{10}$, which is corresponded to $10^{-8}{\rm Hz}<k_{\GW}^{\peak}<10^4{\rm Hz}$. 
 Three examples of $\Omega_{\GW}$ predicted in our model are shown in Fig.~\ref{fig:karidata}. 
 One can see that the GWs sourced by the helical magnetic fields can be observed by the future GW interferometers, DECIGO, BBO and LISA. 

 It is interesting to relate the maximum value of $\Omega_{\GW}$ and  $B_{\rm eff}$. By using Eq.~\eqref{eq:estimation of power spectra} and Eq.~\eqref{eq:beffective}, one can derive their relation as,
 \begin{align}
 \label{eq:BeffandGW}
 B_{\rm eff}\approx 1.6\times 10^{-14}\G\left(\frac{\xi}{10}\right)\left(\frac{\Omega_{\GW}^{\max}}{10^{-13}}\right)^{1/4}\left(\frac{k_{\GW}^{\peak}}{\rm 1Hz}\right)^{-1/2},
 \end{align}
where $\Omega_{\GW}^{\max}$ denotes the maximum value of $\Omega_{\GW}(k)$. 
This relation offers us the comprehensive way to test the prediction of the model
by combining the observations of GWs and IGMFs.
\section{Summary and Discussion}
\label{sec:summary}

In this paper, we have shown that  the helical magnetic fields generated in the hybrid magnetogenesis model proposed in Ref.~\cite{Fujita:2019pmi} can source GWs 
which will be observed by the upcoming GW interferometers.
The peak frequency of the sourced GWs depends on the reheating temperature,
and it comes to the best sensitivity region of LISA and DECIGO (BBO) for $T_r\sim 10^2$GeV and $10^{4}$GeV, respectively.
The GW amplitude at the peak scale is determined by the energy fraction of the electromagnetic fields at the reheating completion, and for $\Omega_{\EM}(\eta_r) \gtrsim 10^{-2}$ and $10^{-4}$ the GW amplitude exceed the sensitivity curves of LISA and DECIGO, respectively.
The power spectra of the generated magnetic fields have the significant peak on the horizon scale at the reheating completion which makes the resultant magnetic fields strong enough to explain the observational lower bound Eq.~\eqref{eq:observation bound}  with the aid of the inverse cascade process.
The contribution from the peak to the effective magnetic strength in Eq.~\eqref{eq:observation bound} was not dominant in the previous work Ref.~\cite{Fujita:2019pmi} because of the different parameter choice.
Since the sourced GWs are maximally helical, in principle, we can observationally distinguish them from the other signals. 
Furthermore, based on Eq.~\eqref{eq:BeffandGW} , once the sourced GWs are observed, 
the effective strength of the magnetic fields $B_{\rm eff}$ is inferred in this model.
Thus the predictions of our model can be verified by the future observations of GWs and cosmic magnetic fields.

Ref.~\cite{Caprini:2014mja} also studied the induced GWs in the original hybrid magnetogenesis model, and the scale-invariant power spectrum of the GWs was obtained.
Since the original model considers magnetogenesis only during the inflation, the significant amplification of the electromagnetic fields due to the tachyonic instability takes place  only when the modes exit the horizon, and the modes decrease on super-horizon scales. 
On the other hand, since our model considers magnetogenesis during inflation and the reheating era, the second tachyonic amplification occurs when the modes re-enter
the horizon. Moreover, the super-horizon modes are increased by the kinetic coupling and the sub-horizon modes quickly decay during reheating.
Putting them altogether, one finds that the electromagnetic spectra in our model acquire significant peaks on the horizon scale during reheating.
As a result, the induced GWs has the significant peak on the horizon scale at the end of magnetogenesis, which provides a fascinating observational signature for the GW interferometers. 

In this paper, we implicitly assumed that the electric part of the generated electromagnetic waves is instantly dissipated at the reheating completion. Then the electromagnetic waves are converted into frozen magnetic fields right after reheating. This assumption is often made in many magnetogenesis works for simplicity. 
However, the results may significantly alter, when we consider another reheating process. If the oscillating inflaton gradually decays into charged particle, for instance,
the electric conductivity induced by the charged particles stops megnetogenesis before the reheating completion as well as the magnetic fields are merely diluted by the cosmic expansion until the inverse cascade begins. In this case, the sourced GWs may be also suppressed, because the GWs undergo the reheating era without the source effect.
We will explore how the predictions of magnetogeneis models change depending on reheating scenarios in the future work.

We did not consider some potentially important phenomena related to the primordial magnetogenesis, such as Schwinger effect, chiral anomaly and baryogenesis in this paper. Schwinger effect is the non-perturbative  phenomenon in the QED, in which the charged particle and anti-particle are generated by the strong electric fields. Since they can induce the electric conductivity, Schwinger effect can affect the dynamics of magnetogenesis~\cite{Schwinger:1951nm,Frob:2014zka,Kobayashi:2014zza,Stahl:2015gaa,Hayashinaka:2016qqn,Bavarsad:2016cxh,Kitamoto:2018htg,Sobol:2018djj,Banyeres:2018aax,Stahl:2018idd,Domcke:2019qmm,Shtanov:2020gjp}. The chiral anomaly associates the helicity of the electromagnetic fields with the chiral asymmetry of the charged fermions which we neglect in this paper. The chiral asymmetry modifies the inverse cascade through the chiral magnetic effect and the final magnetic fields would be significantly different~\cite{Fukushima:2008xe,Boyarsky:2011uy,Akamatsu:2013pjd,Schober:2017cdw,Schober:2018ojn,Domcke:2018eki,Adshead:2018oaa,Domcke:2019qmm,Schober:2020ogz}.
It has been pointed out that helical primordial magnetic fields can be responsible for the generation of the baryon asymmetry in our universe~\cite{Fujita:2016igl,Kamada:2016eeb,Kamada:2016cnb,Barrie:2020kpt}. This mechanism may also constrain magnetogenesis models when the over-production of baryons is predicted. However, the chiral plasma insability may cancel the helicity of the magnetic fields and the over-production of the baryons may be suppressed~\cite{Domcke:2019mnd}. Careful calculations are needed to determine the amount of the cancellation, and it is beyond the scope of this paper.
The above phenomena require dedicated investigations to evaluate their implication and 
it is worth studying them in broader contexts than primordial magnetogenesis.

\acknowledgments
We would like to thank Kohei Kamada, Rampei Kimura, Yoh Kobayashi, Sachiko Kuroyanagi, Kei-ichi Maeda, Shoichiro Miyashita, Priti Gupta, Seiga Sato, Shintaro Sato, Teruaki Suyama, and Masahide Yamaguchi for useful discussions and comments. 
The work of T.F. was supported by JSPS KAKENHI No.~17J09103 and No.~18K13537.

\appendix
\section{The analytic calculation of the $z$-integral}
\label{sec:Appendix}

\begin{figure}
        \centering
        \includegraphics[width=.6\textwidth]{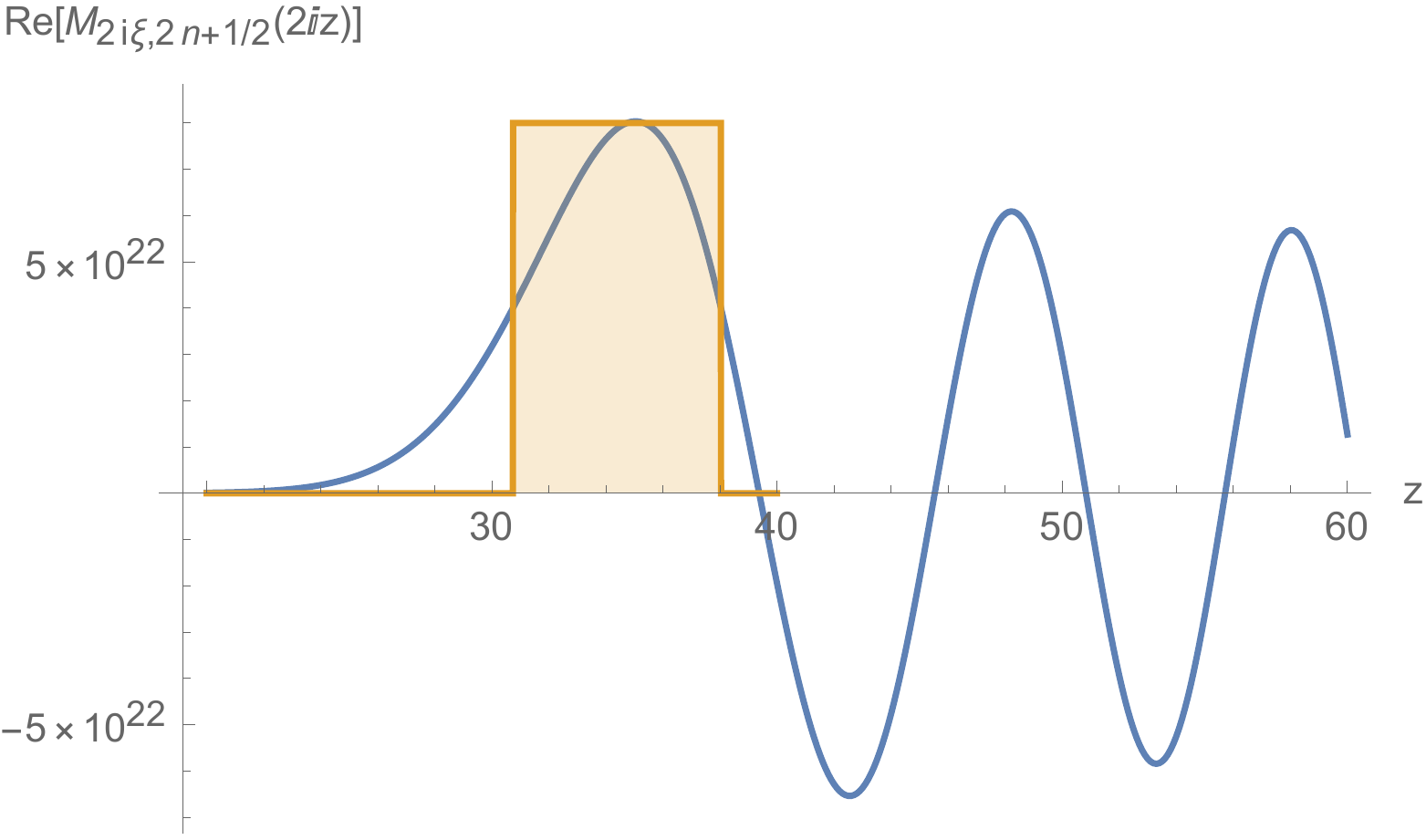}
        \caption{The real part of the Whittaker function, $M_{2i\xi,2n+\frac{1}{2}}(2iz)$. We chose the parameters as $(n,\xi)=(3, 7.6)$. It behaves as the polynomial of z before the first peak at $z\simeq 35$, but later it shows a damping oscillation. The orange box represents the tophat function defined in Eq.~\eqref{eq:tophat approximation} with the peak amplitude and FWHM of $M_{2i\xi,2n+\frac{1}{2}}(2iz)$.}
        \label{fig:WhittakerM}
\end{figure}

\begin{figure}
        \centering
        \includegraphics[width=\textwidth]{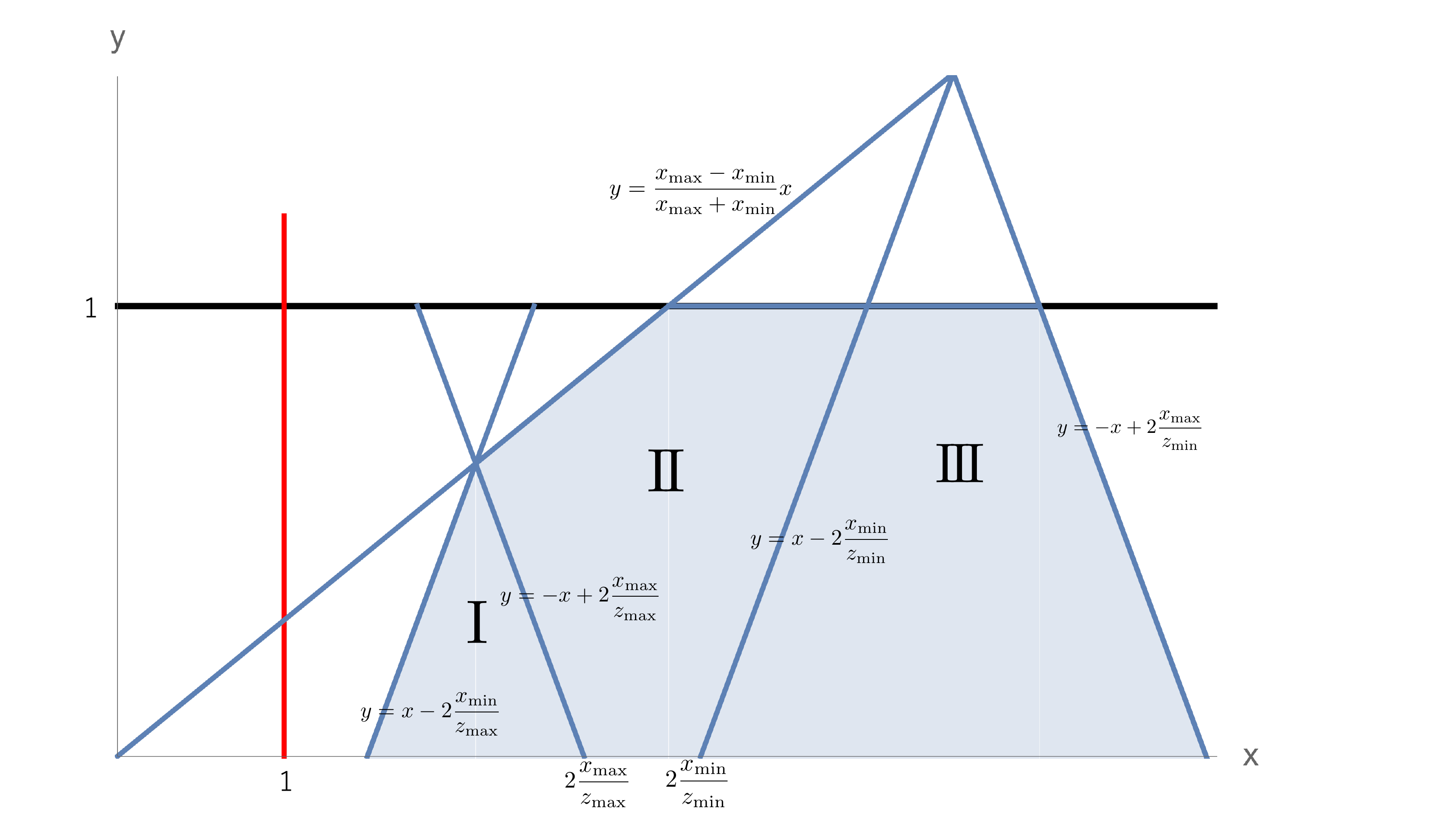}
        \caption{A schematic picture of the integration domain of the momentum integrals over $x=[1,\infty]$ and $y=[0,1]$ in Eq.~\eqref{eq:horn ps}. The red and black solid lines denote the integration limits of $x$ and $y$, respectively. This picture represents the case for $k\eta=8\xi$ which corresponds to calculating the GWs at the peak scale. If the wave number of the GWs is much larger than the peak, the region I vanishes. Otherwise, the contribution from the region I is the dominant component of the integral. }
        \label{fig:haniz1}
\end{figure}

In this appendix, we describe how the $z$-integral in Eq.~\eqref{eq:horn ps} is performed
with the approximation Eq.~\eqref{eq:tophat approximation}.
Fig.~\ref{fig:WhittakerM} shows the Whittaker function, $M_{\pm 2i\xi,2n+\frac{1}{2}}(2ix)$
which determines the time evolution of the growing mode of the gauge fields (see Eq.~\eqref{eq:oscillation sol}). 
Since the prefactor of the Whittaker functions in the $z$ integral in the last line of Eq.~\eqref{eq:horn ps} decays in proportion to $z^{-2}$ for $z\gg1$, the first peak
of  the Whittaker function is expected to have the dominant contribution to the $z$-integral. Thus we can use the approximation, Eq.~\eqref{eq:tophat approximation}.
If we replace $M_{2i\xi,2n+\frac{1}{2}}(2ix)$ with $f_{\rm tophat}(x)$, the z-integral of Eq.~\eqref{eq:horn ps} is rewritten as,
\begin{align}
\int_{z_{\min}}^{z_{\max}}dz G(z)f_{\rm tophat}(\frac{x-y}{2}z)f_{\rm tophat}(\frac{x+y}{2}z),
\end{align}
where we introduced $z_{\max}\equiv k\eta, z_{\min}\equiv 2|k\eta_e|$, and $G(z)\equiv\left(k\eta z +1\right)\sin(k\eta-z)/z^3+(z-k\eta)\cos(k\eta-z)/z^3$. We define the variables as
\begin{align}
r\equiv\frac{x+y}{2}, \quad s\equiv\frac{x-y}{2}, \quad (r>s).
\end{align}
Then, the z-integration is analytically performed  as
\begin{align}
\int_{z_{\min}}^{z_{\max}}dz G(z)f_{\rm tophat}(s z)f_{\rm tophat}(r z)=a^2\left[g(z)\right]_{\max[z_{\min},\frac{x_{\min}}{s}]}^{\min[z_{\max},\frac{x_{\max}}{r}]},
\label{z integrated}
\end{align}
where
\begin{align}
g(z)\equiv &\frac{(k\eta-z)\cos(k\eta-z)-(1+k\eta z)\sin(k\eta-z)}{2z^2}\notag\\
&+\frac{\rm{Ci}(z)(\sin(k\eta)-k\eta\cos(k\eta))-(\cos(k\eta)+k\eta\sin(k\eta)\rm{Si}(z))}{2}.
\end{align}
The result of the $z$-integral, Eq.~\eqref{z integrated}, is the part of the integrand of
the $x$ and $y$ integrals in Eq.~\eqref{eq:horn ps}.
However, the concrete expression of Eq.~\eqref{z integrated} changes depending on the values of $x$ and $y$, and thus one needs to decompose the integration domain into three regions in which 
Eq.~\eqref{z integrated} takes a definite expression. 
These regions are illustrated in Fig.~\ref{fig:haniz1}.
\subsection*{Region I ($r<\frac{x_{\max}}{z_{\max}}\quad\&\quad  s<\frac{x_{\min}}{z_{\min}}$)} 
This region includes the convolution of the peak scale magnetic fields at $\eta$. Then the contribution from this region is dominant. Integration domain of the momentum integration, $x,y$ is written as,
\begin{align}
0<&y<\min[1,-x+2\frac{x_{\max}}{z_{\max}},x-2\frac{x_{\min}}{z_{\max}}],\notag\\
\max[1,2\frac{x_{\min}}{z_{\max}}]<&x<\max[1,2\frac{x_{\max}}{z_{\max}}].
\end{align}
\subsection*{Region I\hspace{-.1em}I ($r>\frac{x_{\max}}{z_{\max}}\quad\&\quad s<\frac{x_{\min}}{z_{\min}}\quad\&\quad \frac{x_{\max}}{r}>\frac{x_{\min}}{s}$)}
This region is not  the dominant component since this region does not include the contribution at $\eta$, which is the upper end of the time integration. The integration domain of $x,y$ is written as
\begin{align}
\max[0,x-2\frac{x_{\min}}{z_{\min}},-x+2\frac{x_{\max}}{z_{\max}}]<&y<\min[1,\frac{x_{\max}-x_{\min}}{x_{\min}+x_{\max}}x],\notag\\
\max[2\frac{x_{\max}}{z_{\max}}-1,1,\frac{x_{\max}+x_{\min}}{z_{\max}}]<&x<\min[2\frac{x_{\min}}{z_{\min}}+1,\frac{x_{\min}+x_{\max}}{z_{\min}}].
\end{align}
\subsection*{Region I\hspace{-.1em}I\hspace{-.1em}I($r>\frac{x_{\max}}{z_{\max}}\quad\&\quad s>\frac{x_{\min}}{z_{\min}}$)}
 This region gives the smallest contribution since this region include the only the convolution of the peak magnetic fields at the inflation end, $2|\eta_e|$. The integration domain for $x,y$ is represented as,
 \begin{align}
 0<&y<\min[1,x-2\frac{x_{\min}}{z_{\min}},-x+2\frac{x_{\max}}{z_{\min}}]\notag\\
 2\frac{x_{\min}}{z_{\min}}<&x<2\frac{x_{\max}}{z_{\min}}
 \end{align}



\begin{thebibliography}{99}
\bibitem{Bernet:2008qp} 
  M.~L.~Bernet, F.~Miniati, S.~J.~Lilly, P.~P.~Kronberg and M.~Dessauges-Zavadsky,
  Nature {\bf 454}, 302 (2008)
  doi:10.1038/nature07105
  [arXiv:0807.3347 [astro-ph]].


\bibitem{Bonafede:2010xg} 
  A.~Bonafede, L.~Feretti, M.~Murgia, F.~Govoni, G.~Giovannini, D.~Dallacasa, K.~Dolag and G.~B.~Taylor,
  Astron.\ Astrophys.\  {\bf 513}, A30 (2010)
  doi:10.1051/0004-6361/200913696
  [arXiv:1002.0594 [astro-ph.CO]].


\bibitem{Feretti:2012vk} 
  L.~Feretti, G.~Giovannini, F.~Govoni and M.~Murgia,
  Astron.\ Astrophys.\ Rev.\  {\bf 20}, 54 (2012)
  doi:10.1007/s00159-012-0054-z
  [arXiv:1205.1919 [astro-ph.CO]].


\bibitem{Neronov:1900zz} 
  A.~Neronov and I.~Vovk,
  Science {\bf 328}, 73 (2010)
  doi:10.1126/science.1184192
  [arXiv:1006.3504 [astro-ph.HE]].


\bibitem{Dolag:2010ni} 
  K.~Dolag, M.~Kachelriess, S.~Ostapchenko and R.~Tomas,
  Astrophys.\ J.\ Lett.\  {\bf 727}, L4 (2011)
  doi:10.1088/2041-8205/727/1/L4
  [arXiv:1009.1782 [astro-ph.HE]].


\bibitem{Essey:2010nd} 
  W.~Essey, S.~Ando and A.~Kusenko,
  Astropart.\ Phys.\  {\bf 35}, 135 (2011)
  doi:10.1016/j.astropartphys.2011.06.010
  [arXiv:1012.5313 [astro-ph.HE]].


\bibitem{Tavecchio:2010ja} 
  F.~Tavecchio, G.~Ghisellini, G.~Bonnoli and L.~Foschini,
  Mon.\ Not.\ Roy.\ Astron.\ Soc.\  {\bf 414}, 3566 (2011)
  doi:10.1111/j.1365-2966.2011.18657.x
  [arXiv:1009.1048 [astro-ph.HE]].


\bibitem{Taylor:2011bn} 
  A.~M.~Taylor, I.~Vovk and A.~Neronov,
  Astron.\ Astrophys.\  {\bf 529}, A144 (2011)
  doi:10.1051/0004-6361/201116441
  [arXiv:1101.0932 [astro-ph.HE]].


\bibitem{Vovk:2011aa} 
  I.~Vovk, A.~M.~Taylor, D.~Semikoz and A.~Neronov,
  Astrophys.\ J.\ Lett.\  {\bf 747}, L14 (2012)
  doi:10.1088/2041-8205/747/1/L14
  [arXiv:1112.2534 [astro-ph.CO]].


\bibitem{Takahashi:2013lba} 
  K.~Takahashi, M.~Mori, K.~Ichiki, S.~Inoue and H.~Takami,
  Astrophys.\ J.\ Lett.\  {\bf 771}, L42 (2013)
  doi:10.1088/2041-8205/771/2/L42
  [arXiv:1303.3069 [astro-ph.CO]].


\bibitem{Chen:2014rsa} 
  W.~Chen, J.~H.~Buckley and F.~Ferrer,
  Phys.\ Rev.\ Lett.\  {\bf 115}, 211103 (2015)
  doi:10.1103/PhysRevLett.115.211103
  [arXiv:1410.7717 [astro-ph.HE]].


\bibitem{Finke:2015ona} 
  J.~D.~Finke, L.~C.~Reyes, M.~Georganopoulos, K.~Reynolds, M.~Ajello, S.~J.~Fegan and K.~McCann,
  Astrophys.\ J.\  {\bf 814}, no. 1, 20 (2015)
  doi:10.1088/0004-637X/814/1/20
  [arXiv:1510.02485 [astro-ph.HE]].


\bibitem{Biteau:2018tmv} 
  M.~Ackermann {\it et al.} [Fermi-LAT Collaboration],
  Astrophys.\ J.\ Suppl.\  {\bf 237}, no. 2, 32 (2018)
  doi:10.3847/1538-4365/aacdf7
  [arXiv:1804.08035 [astro-ph.HE]].


\bibitem{Ade:2015cva} 
  P.~A.~R.~Ade {\it et al.} [Planck Collaboration],
  Astron.\ Astrophys.\  {\bf 594}, A19 (2016)
  doi:10.1051/0004-6361/201525821
  [arXiv:1502.01594 [astro-ph.CO]].


\bibitem{Durrer:2013pga} 
  R.~Durrer and A.~Neronov,
  Astron.\ Astrophys.\ Rev.\  {\bf 21}, 62 (2013)
  doi:10.1007/s00159-013-0062-7
  [arXiv:1303.7121 [astro-ph.CO]].


\bibitem{Subramanian:2015lua} 
  K.~Subramanian,
  Rept.\ Prog.\ Phys.\  {\bf 79}, no. 7, 076901 (2016)
  doi:10.1088/0034-4885/79/7/076901
  [arXiv:1504.02311 [astro-ph.CO]].


\bibitem{Hanayama:2005hd} 
  H.~Hanayama, K.~Takahashi, K.~Kotake, M.~Oguri, K.~Ichiki and H.~Ohno,
  Astrophys.\ J.\  {\bf 633}, 941 (2005)
  doi:10.1086/491575
  [astro-ph/0501538].


\bibitem{Davis:1999bt} 
  A.~C.~Davis, M.~Lilley and O.~Tornkvist,
  Phys.\ Rev.\ D {\bf 60}, 021301 (1999)
  doi:10.1103/PhysRevD.60.021301
  [astro-ph/9904022].


\bibitem{Vachaspati:1991nm} 
  T.~Vachaspati,
  Phys.\ Lett.\ B {\bf 265}, 258 (1991).
  doi:10.1016/0370-2693(91)90051-Q


\bibitem{Enqvist:1993np} 
  K.~Enqvist and P.~Olesen,
  Phys.\ Lett.\ B {\bf 319}, 178 (1993)
  doi:10.1016/0370-2693(93)90799-N
  [hep-ph/9308270].


\bibitem{Grasso:1997nx} 
  D.~Grasso and A.~Riotto,
  Phys.\ Lett.\ B {\bf 418}, 258 (1998)
  doi:10.1016/S0370-2693(97)01224-0
  [hep-ph/9707265].


\bibitem{Ellis:2019tjf} 
  J.~Ellis, M.~Fairbairn, M.~Lewicki, V.~Vaskonen and A.~Wickens,
  JCAP {\bf 1909}, 019 (2019)
  doi:10.1088/1475-7516/2019/09/019
  [arXiv:1907.04315 [astro-ph.CO]].


\bibitem{Takahashi:2005nd} 
  K.~Takahashi, K.~Ichiki, H.~Ohno and H.~Hanayama,
  Phys.\ Rev.\ Lett.\  {\bf 95}, 121301 (2005)
  doi:10.1103/PhysRevLett.95.121301
  [astro-ph/0502283].


\bibitem{Saga:2015bna} 
  S.~Saga, K.~Ichiki, K.~Takahashi and N.~Sugiyama,
  Phys.\ Rev.\ D {\bf 91}, no. 12, 123510 (2015)
  doi:10.1103/PhysRevD.91.123510
  [arXiv:1504.03790 [astro-ph.CO]].


\bibitem{Benevides:2018mwx} 
  A.~Benevides, A.~Dabholkar and T.~Kobayashi,
  JHEP {\bf 1811}, 039 (2018)
  doi:10.1007/JHEP11(2018)039
  [arXiv:1808.08237 [hep-th]].


\bibitem{Ratra:1991bn} 
  B.~Ratra,
  Astrophys.\ J.\ Lett.\  {\bf 391}, L1 (1992).
  doi:10.1086/186384


\bibitem{Bamba:2003av} 
  K.~Bamba and J.~Yokoyama,
  Phys.\ Rev.\ D {\bf 69}, 043507 (2004)
  doi:10.1103/PhysRevD.69.043507
  [astro-ph/0310824].


\bibitem{Demozzi:2009fu} 
  V.~Demozzi, V.~Mukhanov and H.~Rubinstein,
  JCAP {\bf 0908}, 025 (2009)
  doi:10.1088/1475-7516/2009/08/025
  [arXiv:0907.1030 [astro-ph.CO]].


\bibitem{Fujita:2012rb} 
  T.~Fujita and S.~Mukohyama,
  JCAP {\bf 1210}, 034 (2012)
  doi:10.1088/1475-7516/2012/10/034
  [arXiv:1205.5031 [astro-ph.CO]].


\bibitem{Fujita:2013qxa} 
  T.~Fujita and S.~Yokoyama,
  JCAP {\bf 1309}, 009 (2013)
  doi:10.1088/1475-7516/2013/09/009
  [arXiv:1306.2992 [astro-ph.CO]].


\bibitem{Fujita:2014sna} 
  T.~Fujita and S.~Yokoyama,
  JCAP {\bf 1403}, 013 (2014)
  Erratum: [JCAP {\bf 1405}, E02 (2014)]
  doi:10.1088/1475-7516/2014/03/013, 10.1088/1475-7516/2014/05/E02
  [arXiv:1402.0596 [astro-ph.CO]].


\bibitem{Ferreira:2013sqa} 
  R.~J.~Z.~Ferreira, R.~K.~Jain and M.~S.~Sloth,
  JCAP {\bf 1310}, 004 (2013)
  doi:10.1088/1475-7516/2013/10/004
  [arXiv:1305.7151 [astro-ph.CO]].


\bibitem{Ferreira:2014hma} 
  R.~J.~Z.~Ferreira, R.~K.~Jain and M.~S.~Sloth,
  JCAP {\bf 1406}, 053 (2014)
  doi:10.1088/1475-7516/2014/06/053
  [arXiv:1403.5516 [astro-ph.CO]].


\bibitem{Kobayashi:2014sga} 
  T.~Kobayashi,
  JCAP {\bf 1405}, 040 (2014)
  doi:10.1088/1475-7516/2014/05/040
  [arXiv:1403.5168 [astro-ph.CO]].


\bibitem{Fujita:2016qab} 
  T.~Fujita and R.~Namba,
  Phys.\ Rev.\ D {\bf 94}, no. 4, 043523 (2016)
  doi:10.1103/PhysRevD.94.043523
  [arXiv:1602.05673 [astro-ph.CO]].


\bibitem{Vilchinskii:2017qul} 
  S.~Vilchinskii, O.~Sobol, E.~Gorbar and I.~Rudenok,
  Phys.\ Rev.\ D {\bf 95}, no. 8, 083509 (2017)
  doi:10.1103/PhysRevD.95.083509
  [arXiv:1702.02774 [astro-ph.CO]].


\bibitem{Turner:1987bw} 
  M.~S.~Turner and L.~M.~Widrow,
  Phys.\ Rev.\ D {\bf 37}, 2743 (1988).
  doi:10.1103/PhysRevD.37.2743


\bibitem{Garretson:1992vt} 
  W.~D.~Garretson, G.~B.~Field and S.~M.~Carroll,
  Phys.\ Rev.\ D {\bf 46}, 5346 (1992)
  doi:10.1103/PhysRevD.46.5346
  [hep-ph/9209238].


\bibitem{Field:1998hi} 
  G.~B.~Field and S.~M.~Carroll,
  Phys.\ Rev.\ D {\bf 62}, 103008 (2000)
  doi:10.1103/PhysRevD.62.103008
  [astro-ph/9811206].


\bibitem{Anber:2006xt} 
  M.~M.~Anber and L.~Sorbo,
  JCAP {\bf 0610}, 018 (2006)
  doi:10.1088/1475-7516/2006/10/018
  [astro-ph/0606534].


\bibitem{Fujita:2015iga} 
  T.~Fujita, R.~Namba, Y.~Tada, N.~Takeda and H.~Tashiro,
  JCAP {\bf 1505}, 054 (2015)
  doi:10.1088/1475-7516/2015/05/054
  [arXiv:1503.05802 [astro-ph.CO]].


\bibitem{Adshead:2016iae} 
  P.~Adshead, J.~T.~Giblin, T.~R.~Scully and E.~I.~Sfakianakis,
  JCAP {\bf 1610}, 039 (2016)
  doi:10.1088/1475-7516/2016/10/039
  [arXiv:1606.08474 [astro-ph.CO]].


\bibitem{Caprini:2014mja} 
  C.~Caprini and L.~Sorbo,
  JCAP {\bf 1410}, 056 (2014)
  doi:10.1088/1475-7516/2014/10/056
  [arXiv:1407.2809 [astro-ph.CO]].


\bibitem{Caprini:2017vnn} 
  C.~Caprini, M.~C.~Guzzetti and L.~Sorbo,
  Class.\ Quant.\ Grav.\  {\bf 35}, no. 12, 124003 (2018)
  doi:10.1088/1361-6382/aac143
  [arXiv:1707.09750 [astro-ph.CO]].


\bibitem{Son:1998my} 
  D.~T.~Son,
  Phys.\ Rev.\ D {\bf 59}, 063008 (1999)
  doi:10.1103/PhysRevD.59.063008
  [hep-ph/9803412].


\bibitem{Christensson:2000sp} 
  M.~Christensson, M.~Hindmarsh and A.~Brandenburg,
  Phys.\ Rev.\ E {\bf 64}, 056405 (2001)
  doi:10.1103/PhysRevE.64.056405
  [astro-ph/0011321].


\bibitem{Kahniashvili:2012uj} 
  T.~Kahniashvili, A.~G.~Tevzadze, A.~Brandenburg and A.~Neronov,
  Phys.\ Rev.\ D {\bf 87}, no. 8, 083007 (2013)
  doi:10.1103/PhysRevD.87.083007
  [arXiv:1212.0596 [astro-ph.CO]].


\bibitem{Sharma:2018kgs} 
  R.~Sharma, K.~Subramanian and T.~R.~Seshadri,
  Phys.\ Rev.\ D {\bf 97}, no. 8, 083503 (2018)
  doi:10.1103/PhysRevD.97.083503
  [arXiv:1802.04847 [astro-ph.CO]].


\bibitem{Fujita:2019pmi} 
  T.~Fujita and R.~Durrer,
  JCAP {\bf 1909}, 008 (2019)
  doi:10.1088/1475-7516/2019/09/008
  [arXiv:1904.11428 [astro-ph.CO]].


\bibitem{Domenech:2015zzi} 
  G.~Domènech, C.~Lin and M.~Sasaki,
  EPL {\bf 115}, no. 1, 19001 (2016)
  doi:10.1209/0295-5075/115/19001
  [arXiv:1512.01108 [astro-ph.CO]].


\bibitem{Mukohyama:2016npi} 
  S.~Mukohyama,
  Phys.\ Rev.\ D {\bf 94}, no. 12, 121302 (2016)
  doi:10.1103/PhysRevD.94.121302
  [arXiv:1607.07041 [hep-th]].


\bibitem{Brandenburg:2020vwp} 
  A.~Brandenburg, R.~Durrer, Y.~Huang, T.~Kahniashvili, S.~Mandal and S.~Mukohyama,
  arXiv:2005.06449 [astro-ph.CO].


\bibitem{Sorbo:2011rz} 
  L.~Sorbo,
  JCAP {\bf 1106}, 003 (2011)
  doi:10.1088/1475-7516/2011/06/003
  [arXiv:1101.1525 [astro-ph.CO]].


\bibitem{Namba:2015gja} 
  R.~Namba, M.~Peloso, M.~Shiraishi, L.~Sorbo and C.~Unal,
  JCAP {\bf 1601}, 041 (2016)
  doi:10.1088/1475-7516/2016/01/041
  [arXiv:1509.07521 [astro-ph.CO]].


\bibitem{Jimenez:2017cdr} 
  D.~Jiménez, K.~Kamada, K.~Schmitz and X.~J.~Xu,
  JCAP {\bf 1712}, 011 (2017)
  doi:10.1088/1475-7516/2017/12/011
  [arXiv:1707.07943 [hep-ph]].


\bibitem{Saga:2018ont} 
  S.~Saga, H.~Tashiro and S.~Yokoyama,
  Phys.\ Rev.\ D {\bf 98}, no. 8, 083518 (2018)
  doi:10.1103/PhysRevD.98.083518
  [arXiv:1807.00561 [astro-ph.CO]].


\bibitem{Sharma:2019jtb} 
  R.~Sharma, K.~Subramanian and T.~R.~Seshadri,
  Phys.\ Rev.\ D {\bf 101}, no. 10, 103526 (2020)
  doi:10.1103/PhysRevD.101.103526
  [arXiv:1912.12089 [astro-ph.CO]].


\bibitem{Ozsoy:2020ccy} 
  O.~Özsoy,
  arXiv:2005.10280 [astro-ph.CO].


\bibitem{Nakayama:2008wy} 
  K.~Nakayama, S.~Saito, Y.~Suwa and J.~Yokoyama,
  JCAP {\bf 0806}, 020 (2008)
  doi:10.1088/1475-7516/2008/06/020
  [arXiv:0804.1827 [astro-ph]].


\bibitem{Schwinger:1951nm} 
  J.~S.~Schwinger,
  Phys.\ Rev.\  {\bf 82}, 664 (1951).
  doi:10.1103/PhysRev.82.664


\bibitem{Frob:2014zka} 
  M.~B.~Fröb, J.~Garriga, S.~Kanno, M.~Sasaki, J.~Soda, T.~Tanaka and A.~Vilenkin,
  JCAP {\bf 1404}, 009 (2014)
  doi:10.1088/1475-7516/2014/04/009
  [arXiv:1401.4137 [hep-th]].


\bibitem{Kobayashi:2014zza} 
  T.~Kobayashi and N.~Afshordi,
  JHEP {\bf 1410}, 166 (2014)
  doi:10.1007/JHEP10(2014)166
  [arXiv:1408.4141 [hep-th]].


\bibitem{Stahl:2015gaa} 
  C.~Stahl, E.~Strobel and S.~S.~Xue,
  Phys.\ Rev.\ D {\bf 93}, no. 2, 025004 (2016)
  doi:10.1103/PhysRevD.93.025004
  [arXiv:1507.01686 [gr-qc]].


\bibitem{Hayashinaka:2016qqn} 
  T.~Hayashinaka, T.~Fujita and J.~Yokoyama,
  JCAP {\bf 1607}, 010 (2016)
  doi:10.1088/1475-7516/2016/07/010
  [arXiv:1603.04165 [hep-th]].


\bibitem{Bavarsad:2016cxh} 
  E.~Bavarsad, C.~Stahl and S.~S.~Xue,
  Phys.\ Rev.\ D {\bf 94}, no. 10, 104011 (2016)
  doi:10.1103/PhysRevD.94.104011
  [arXiv:1602.06556 [hep-th]].


\bibitem{Kitamoto:2018htg} 
  H.~Kitamoto,
  Phys.\ Rev.\ D {\bf 98}, no. 10, 103512 (2018)
  doi:10.1103/PhysRevD.98.103512
  [arXiv:1807.03753 [hep-th]].


\bibitem{Sobol:2018djj} 
  O.~O.~Sobol, E.~V.~Gorbar, M.~Kamarpour and S.~I.~Vilchinskii,
  Phys.\ Rev.\ D {\bf 98}, no. 6, 063534 (2018)
  doi:10.1103/PhysRevD.98.063534
  [arXiv:1807.09851 [hep-ph]].


\bibitem{Banyeres:2018aax} 
  M.~Banyeres, G.~Domènech and J.~Garriga,
  JCAP {\bf 1810}, 023 (2018)
  doi:10.1088/1475-7516/2018/10/023
  [arXiv:1809.08977 [hep-th]].


\bibitem{Stahl:2018idd} 
  C.~Stahl,
  Nucl.\ Phys.\ B {\bf 939}, 95 (2019)
  doi:10.1016/j.nuclphysb.2018.12.017
  [arXiv:1806.06692 [hep-th]].


\bibitem{Domcke:2019qmm} 
  V.~Domcke, Y.~Ema and K.~Mukaida,
  JHEP {\bf 2002}, 055 (2020)
  doi:10.1007/JHEP02(2020)055
  [arXiv:1910.01205 [hep-ph]].


\bibitem{Shtanov:2020gjp} 
  Y.~Shtanov and M.~Pavliuk,
  arXiv:2004.00947 [astro-ph.CO].


\bibitem{Fukushima:2008xe} 
  K.~Fukushima, D.~E.~Kharzeev and H.~J.~Warringa,
  Phys.\ Rev.\ D {\bf 78}, 074033 (2008)
  doi:10.1103/PhysRevD.78.074033
  [arXiv:0808.3382 [hep-ph]].


\bibitem{Boyarsky:2011uy} 
  A.~Boyarsky, J.~Frohlich and O.~Ruchayskiy,
  Phys.\ Rev.\ Lett.\  {\bf 108}, 031301 (2012)
  doi:10.1103/PhysRevLett.108.031301
  [arXiv:1109.3350 [astro-ph.CO]].


\bibitem{Akamatsu:2013pjd} 
  Y.~Akamatsu and N.~Yamamoto,
  Phys.\ Rev.\ Lett.\  {\bf 111}, 052002 (2013)
  doi:10.1103/PhysRevLett.111.052002
  [arXiv:1302.2125 [nucl-th]].


\bibitem{Schober:2017cdw} 
  J.~Schober, I.~Rogachevskii, A.~Brandenburg, A.~Boyarsky, J.~Fröhlich, O.~Ruchayskiy and N.~Kleeorin,
  Astrophys.\ J.\  {\bf 858}, no. 2, 124 (2018)
  doi:10.3847/1538-4357/aaba75
  [arXiv:1711.09733 [physics.flu-dyn]].


\bibitem{Schober:2018ojn} 
  J.~Schober, A.~Brandenburg, I.~Rogachevskii and N.~Kleeorin,
  Geophys.\ Astrophys.\ Fluid Dynamics {\bf 113}, no. 1-2, 107 (2019)
  doi:10.1080/03091929.2018.1515313
  [arXiv:1803.06350 [physics.flu-dyn]].


\bibitem{Domcke:2018eki} 
  V.~Domcke and K.~Mukaida,
  JCAP {\bf 1811}, 020 (2018)
  doi:10.1088/1475-7516/2018/11/020
  [arXiv:1806.08769 [hep-ph]].


\bibitem{Adshead:2018oaa} 
  P.~Adshead, L.~Pearce, M.~Peloso, M.~A.~Roberts and L.~Sorbo,
  JCAP {\bf 1806}, 020 (2018)
  doi:10.1088/1475-7516/2018/06/020
  [arXiv:1803.04501 [astro-ph.CO]].


\bibitem{Schober:2020ogz} 
  J.~Schober, T.~Fujita and R.~Durrer,
  Phys.\ Rev.\ D {\bf 101}, no. 10, 103028 (2020)
  doi:10.1103/PhysRevD.101.103028
  [arXiv:2002.09501 [physics.plasm-ph]].


\bibitem{Fujita:2016igl} 
  T.~Fujita and K.~Kamada,
  Phys.\ Rev.\ D {\bf 93}, no. 8, 083520 (2016)
  doi:10.1103/PhysRevD.93.083520
  [arXiv:1602.02109 [hep-ph]].


\bibitem{Kamada:2016eeb} 
  K.~Kamada and A.~J.~Long,
  Phys.\ Rev.\ D {\bf 94}, no. 6, 063501 (2016)
  doi:10.1103/PhysRevD.94.063501
  [arXiv:1606.08891 [astro-ph.CO]].


\bibitem{Kamada:2016cnb} 
  K.~Kamada and A.~J.~Long,
  Phys.\ Rev.\ D {\bf 94}, no. 12, 123509 (2016)
  doi:10.1103/PhysRevD.94.123509
  [arXiv:1610.03074 [hep-ph]].


\bibitem{Barrie:2020kpt} 
  N.~D.~Barrie,
  arXiv:2001.04773 [hep-ph].


\bibitem{Domcke:2019mnd} 
  V.~Domcke, B.~von Harling, E.~Morgante and K.~Mukaida,
  JCAP {\bf 1910}, 032 (2019)
  doi:10.1088/1475-7516/2019/10/032
  [arXiv:1905.13318 [hep-ph]].

\end{thebibliography}
\end{document}